\documentclass[twoside,12pt]{article}
\usepackage{epsfig}

\newcommand{\be}{\begin{equation}}
\newcommand{\ee}{\end{equation}}
\newcommand{\bea}{\begin{eqnarray}}
\newcommand{\eea}{\end{eqnarray}}

\usepackage{amsmath}
\usepackage{amsfonts}
\usepackage{amssymb}
\usepackage{amsxtra}
\usepackage{supertabular}
\begin{document}
\title{ \vspace{1cm} 
Do we understand the internal spaces of second quantized fermion and 
boson fields, with gravity included?\\
 Relation with strings theories}

\author{N.S.\ Manko\v c Bor\v stnik$^{1}$,
\\
$^1$Department of Physics, University of Ljubljana\\
SI-1000 Ljubljana, Slovenia}

\maketitle
 

\begin{abstract}
The article proposes the description of internal spaces of 
fermion (quarks and leptons and antiquarks and antileptons)
and boson (photons, weak bosons, gluons, gravitons and scalars) 
second quantized fields in an unique way if they all are 
massless. 
The internal spaces are described by ``basis vectors'' 
which are the superposition of odd (for fermions)
and even (for bosons) products of the operators
$\gamma^ {a}$.
For an arbitrarily symmetry $SO(d-1,1)$ of the 
internal spaces, it is the number of fermion fields (they
appear in families and have their Hermitian conjugated
partners in a separate group) equal to the number of boson 
fields (they appear in two orthogonal groups), manifesting 
a kind of supersymmetry, which differ of the string 
supersymmetry.
On the assumption that fermions and bosons are active (they
have momenta different from zero) only in $d=(3+1)$ ordinary
space-time, bosons present vectors if they carry the space index
$\mu=(0,1,2,3)$, and present scalars if they carry the index
$\sigma \ge 5$.
The author discusses this theory's latest
achievements, with a trial to understand whether the extension
to strings or to odd-dimensional  spaces can lead to new kind of
supersymmetry.
 This model, named {\it spin-charge-family} theory, manifests 
 in a long series of papers ~\cite{norma93,pikanorma2005,nh02,
 n2014matterantimatter,nd2017,%
 2020PartIPartII,nh2021RPPNP,n2023NPB,n2023MDPI,%
 n2024NPB,nh2023dec},  
the phenomenological success of the theory in elementary 
particle physics and cosmology.
%
%
\end{abstract} 

%
\section{Introduction}
\label{introduction}

The proposed article discusses how all the fermion and 
boson fields can be treated uniquely if they all start as massless 
fields. One can, namely, describe the internal spaces of fermions 
and bosons by ``basis vectors'' which are products of
nilpotents and  projectors, chosen to be the eigenvectors of the 
Cartan subalgebra members of the Lorentz algebra in internal 
spaces of fermions and bosons~\cite{norma93,pikanorma2005,%
nh02,n2014matterantimatter,nd2017,2020PartIPartII,
nh2021RPPNP,n2023NPB,n2023MDPI, n2024NPB,nh2023dec}.

Nilpotents are the superposition of an odd number of operators 
$\gamma^{a}$'s ($\stackrel{ab}{(k)}:
=\frac{1}{2} (\gamma^a + \frac{\eta^{aa}}{ik} \gamma^b)\,, 
(\stackrel{ab}{(k)})^2=0$); 
projectors are the superposition of an even number of operators 
$\gamma^{a}$'s ($\stackrel{ab}{[k]}:
= \frac{1}{2}(1+ \frac{i}{k} \gamma^a \gamma^b)\,,
(\stackrel{ab}{[k]})^2=\stackrel{ab}{[k]})$, 
with the properties :
$S^{ab} \,\stackrel{ab}{(k)} = \frac{k}{2} \,\stackrel{ab}{(k)}$\,,
$S^{ab}\,\stackrel{ab}{[k]} = \frac{k}{2} \,\stackrel{ab}{[k]}$\,
with $k^2=\eta^{aa} \eta^{bb}$.

``Basis vectors'' of fermions, chosen to be the algebraic products 
of an odd number of nilpotents (at least one, the rest are 
projectors), and of bosons, chosen to be the algebraic products 
of an even number of nilpotents (or only of projectors) are in even-dimensional spaces correspondingly eigenvectors of all 
$\frac{d}{2}$ (in odd-dimensional spaces of $\frac{ (d-1)}{2}$) Cartan subalgebra members.

Fermion ``basis vectors'' appear in  $2^{\frac{d}{2}-1}$
irreducible representations --- families --- each family having 
$2^{\frac{d}{2}-1}$ members. All the fermion ``basis vectors''  
are mutually orthogonal, while the ``basis vectors'' fulfil
together with their Hermitian conjugated partners, appearing in
a separate group, the Dirac second quantization postulates for 
fermions, therefore, explaining the Dirac's postulates for 
fermions. Fermion ``basis vectors'' and their Hermitian 
conjugated  partners have together $2^{d-1}$ members.
 
Boson ``basis vectors'' appear in two orthogonal groups, each 
of the two groups with $2^{\frac{d}{2}-1}\times$
$2^{\frac{d}{2}-1}$ members have their Hermitian 
conjugated partners within the same group.

Correspondingly, the number of fermion
``basis vectors'' is equal to the number of boson ``basis 
vectors'' manifesting a kind of supersymmetry, which differ
from the one offered by string theories~\footnote{
The supersymmetry usually requires that to each fermion 
there exists a superpartner with the same charges but with 
spin zero; and for each spin 1 boson there exist a superpartner
with the same charges but with spin 1/2. 
}, requiring the 
doubling of so far observed fermions and bosons.

Following the Grassmann algebra, the theory proposes two 
kinds of operators $\gamma^a$, namely $\gamma^a$'s and 
$\tilde{\gamma}^a$'s~\cite{norma93,pikanorma2005,%
nh02,n2014matterantimatter}. 

Operators $\gamma^a$ are used to describe the internal 
spaces of fermions, $\tilde{\gamma}^a$ determine the
quantum numbers of families. The operators $\gamma^a$ 
describe the internal space of bosons as well~\footnote{%
It is not difficult to reproduce the Dirac matrices in any 
$d=2n$ dimensional space, but it is no need for this; the 
``basis vectors'' are much more appropriate to work with, 
what it will be demonstrated in what follows. Let us point 
out that Dirac matrices were designed for massive fermions; 
while ``basis vectors'' describing internal spaces for 
fermions anti-commute, the Dirac matrices do 
not~\cite{tdn2014}. The families do not appear in the
Dirac formulations of internal spaces of fermions.
}.

The operators $\gamma^a$ and $\tilde{\gamma}^a$ 
fulfil the following commutation relations: 
$\{\gamma^{a}, \gamma^{b}\}_{+}=2 \eta^{a b}= 
\{\tilde{\gamma}^{a},
\tilde{\gamma}^{b}\}_{+}$\,,
$\{\gamma^{a}, \tilde{\gamma}^{b}\}_{+}=0\,,
(a,b)=(0,1,2,3,5,\cdots,d)$\,, 
$(\gamma^{a})^{\dagger} = \eta^{aa}\, \gamma^{a}\, ,
(\tilde{\gamma}^{a})^{\dagger} = \eta^{a a}\, 
\tilde{\gamma}^{a}$\,,
making fermions and bosons second quantized fields since 
fermion ``basis vectors'' including odd products of nilpotents 
correspondingly anti-commute and boson ``basis vectors'' 
including even products of nilpotents 
commute~\cite{norma93,pikanorma2005,%
nh02,n2014matterantimatter,nh2021RPPNP,n2023NPB,%
n2023MDPI, n2024NPB}.

Although the multiplication of  the fermion ``basis vectors'' 
by a $\gamma^a$ generates the boson ``basis vectors'', 
and the multiplication of the boson ``basis vectors'' by a 
$\gamma^a$ generates the fermion ``basis vectors'', these 
two kinds of ``basis vectors'' have completely different 
properties: Not only that the fermion  ``basis vectors'' 
anti-commute while the boson ``basis vectors'' commute;
the fermion ``basis vectors'' appear in  $2^{\frac{d}{2}-1}$
families, each family having $2^{\frac{d}{2}-1}$ members,
together with their Hermitian conjugated partners, which appear
in a separate group, they have $2 \times$ $2^{\frac{d}{2}-1}
\times$ $2^{\frac{d}{2}-1}$ members;
the boson ``basis vectors'' appear in two orthogonal groups,
each of the two groups with $2^{\frac{d}{2}-1}\times$
$2^{\frac{d}{2}-1}$ members have their Hermitian conjugated
partners within the same group; the algebraic multiplication of 
one of fermion ``basis vectors'' with one of the Hermitian 
conjugated partner generates  one of the  even ``basis 
vectors''; the algebraic multiplication of one of boson ``basis 
vector'' algebraically applying on a fermion ``basis vector'' 
generates a fermion ``basic vector'', demonstrating that 
since fermion ``basis vectors'' carry a half-integer value of
the Cartan subalgebra members ($S^{ab}=\pm \frac{i}{2}$ 
or $S^{ab}=\pm \frac{1}{2}$, the same is true for 
$\tilde{S}^{ab}=\pm \frac{i}{2}$ or $\tilde{S}^{ab}=
\pm \frac{1}{2}$), the  boson ``basis vectors'' carry an 
integer value of the Cartan subalgebra members, determined
by  ${\cal {\bf S}}^{ab} = S^{ab} +\tilde{S}^{ab}$.

These, explained above, are valid in even dimensional spaces.
In odd dimensional spaces~\cite{n2023MDPI}, $d=(2n+1), 
n\ge 0$, there are two 
different kinds of ``basis vectors'': One kind behaves as they 
do in  $2n$ even dimensional spaces; the second kind with the 
same number of ``basis vectors'' as the first one behaves 
completely different --- the anti commuting ``basis vectors''  
appear in two orthogonal groups, each with their Hermitian
conjugated partners in their group; the commuting
``basis vectors''  appear $2^{\frac{2n}{2}-1}$ families,
each family having $2^{\frac{2n}{2}-1}$ members, while
their Hermitian conjugated partners appear in a separate
group.
The second kind offers the presence of the Fadeev-Popov 
ghosts~\cite{FadeevPopov}.  

The proposed ``basis vectors'', describing the internal spaces 
of fermions and bosons in a tensor product with the basis in 
ordinary space-time determine creation and annihilation 
operators, which explain the Dirac's second quantized 
postulates. 

Under the assumption that fermions and bosons are active 
(have non zero momenta) in $d=(3+1)$ while manifesting 
the internal space in $d=2(2n+1)$ (the experiments and 
observations require $n=3$), the theory offers the 
explanations of all the assumptions of the Standard model 
for fermions (quarks and leptons and antiquarks and 
antileptons), requiring the existence of a right handed 
neutrinos and left handed anti-neutrinos and of families; the 
theory explains the existence of photons, weak bosons and 
gluons, requiring the existence of gravitons. 

The theory explains the existence of scalar fields as it is 
Higgs boson, predicting new scalar fields, which  caused the 
inflation at the Big Bang, explains several cosmological 
observations.

We briefly discuss the case that the point fermion and boson 
fields are extended to strings, describing the fermion and boson 
fields in space-time $d=(3+1)$ as tensor products of the 
``basis vectors'' and basis in ordinary space-time extended 
to strings~\cite{nh2023dec}. This article presents shortly
also a possibility that the internal space of a string, which 
has $d=(1+1)$, Sect.~\ref{string}, is used.
This part is new and yet to be discussed. The longer version 
will be discussed in a separate article in this proceedings.


Sect.~\ref{basisvectors0} presents the properties of ``basis
vectors'' of fermions and bosons. We show how to
construct  the internal spaces of fermions and bosons, and
also demonstrate that the internal spaces of bosons (photons,
gravitons, gauge fields, scalar fields) are expressible as
algebraic products of the fermion ``basis vectors'' and their
Hermitian conjugated partners,
Sect.~\ref{tables},~\cite{n2024NPB}, meaning that we do
not have to know bosons' internal spaces since they can
be presented by algebraic products from fermions' ones.

In Subsect.~\ref{bosons13+1and5+1} ``basis vectors'' for
fermions and bosons in $d=2(2n+1)$-dimensional internal
spaces from the point of view of $d=(3+1)$ are discussed.

Subsect.~\ref{relationsCliffordoddeven}  presents the
relations among the odd and the  even ``basis
vectors''.

Subsect.~\ref{FermionsVetorScalar} discusses relations
among fermions and their vector and scalar gauge fields
under the assumption that all the gauge fields are active
(have non-zero momentum) only in $d=(3+1)$.

Subsect.~\ref{d=(13+1)} discusses the case with
$d=(13+1)$, which explains all the
assumptions of the {\it Standard model},  with the families
included, requires the existence of right-handed
neutrinos and the left-handed antineutrinos, the dark matter,
predicts the fourth family to the observed three, offers the
the explanation for the inflation after Bing Bang and explains also
several cosmological observations.

Sect.~\ref{creationannihilation} presents creation and
annihilation operators for fermion and boson fields
in $d=(3+1)$.

Sect.~\ref{string} discusses the possibility of extending ordinary
space-time to strings, as well as the offer of the extension of an
even dimensional space to one dimension more.

In Sect.~\ref{conclusions}, we shortly overview the whole talk,
pointing out the open questions.

\section{``Basis vectors''  describing internal spaces of anticommuting 
fermion and commuting boson second quantized fields}
\label{basisvectors0}
%
This section is a short overview of the reference~\cite{nh2021RPPNP} 
(and the references~\cite{norma93,nh02}) presenting the ``basis 
vectors'' of fermion internal spaces, while the presentation of the 
``basis vectors'' describing the internal spaces of boson internal spaces 
follows references~\cite{norma93,n2023NPB,n2024NPB}.

We could start with the Grassmann algebra~\cite{norma93,n2023NPB} 
which offers $2\times 2^{d}$ anticommuting operators in 
$d$-dimensional space~\cite{n2019PRD},  $\theta^{a}$ and $p^{\theta a}=
 i \,\frac{\partial}{\partial \theta_{a}}$ ~\cite{norma93}, fulfilling the relations
$\{\theta^{a}, \theta^{b}\}_{+}=0\,,$
$\{\frac{\partial}{\partial \theta_{a}}, 
\frac{\partial}{\partial \theta_{b}}\}_{+} =0\,,$
$\{\theta_{a},\frac{\partial}{\partial \theta_{b}}\}_{+} =\delta_{ab}\,,
\;(a,b)=(0,1,2,3,5,\cdots,d)\,$.

We shall rather use two kinds of the Clifford algebra elements (operators), 
$\gamma^{a}$ and $\tilde{\gamma}^{a}$, expressible with $\theta^{a}$'s 
and their Hermitian conjugate momenta
$p^{\theta a}= i \,\frac{\partial}{\partial \theta_{a}}$ ~\cite{norma93},
\begin{eqnarray}
\label{clifftheta1}
\gamma^{a} &=& (\theta^{a} + \frac{\partial}{\partial \theta_{a}})\,, \quad
\tilde{\gamma}^{a} =i \,(\theta^{a} - \frac{\partial}{\partial \theta_{a}})\,,\nonumber\\
\theta^{a} &=&\frac{1}{2} \,(\gamma^{a} - i \tilde{\gamma}^{a})\,, \quad
\frac{\partial}{\partial \theta_{a}}= \frac{1}{2} \,(\gamma^{a} + i \tilde{\gamma}^{a})\,,
\nonumber\\
\end{eqnarray}
offering together $2\cdot 2^d$ operators: $2^d$ are superposition of 
products of $\gamma^{a}$ and $2^d$ of $\tilde{\gamma}^{a}$,
with the properties
\begin{eqnarray}
\label{gammatildeantiher}
\{\gamma^{a}, \gamma^{b}\}_{+}&=&2 \eta^{a b}= \{\tilde{\gamma}^{a},
\tilde{\gamma}^{b}\}_{+}\,, \nonumber\\
\{\gamma^{a}, \tilde{\gamma}^{b}\}_{+}&=&0\,,\quad
(a,b)=(0,1,2,3,5,\cdots,d)\,, \nonumber\\
(\gamma^{a})^{\dagger} &=& \eta^{aa}\, \gamma^{a}\, , \quad
(\tilde{\gamma}^{a})^{\dagger} = \eta^{a a}\, \tilde{\gamma}^{a}\,.
\end{eqnarray}
Both kinds offer the description of the internal spaces of fermions with the
``basis vectors'' which are superpositions of odd products of either
$\gamma^{a}$'s or $\tilde{\gamma}^{a}$'s and fulfil correspondingly,
the anti-commuting postulates of second quantized fermion fields, as well as
the description of the internal spaces of boson fields with the ``basis
vectors'' which are superposition of even products of either $\gamma^{a}$'s
or $\tilde{\gamma}^{a}$'s and fulfil correspondingly the commuting
postulates of second quantized boson fields. 

{\it Since there are not two kinds of anti-commuting fermions, and not two
corresponding kinds of their gauge fields, the postulate of
Eq.~(\ref{tildegammareduced0}) gives the possibility that only one of the
two kinds of operators are used to describe fermions and their gauge fields,
namely} $\gamma^{a}$'s.\\

 {\it Postulating} how does $\tilde{\gamma}^{a}$ operate on $\gamma^a$, 
reduces the two Clifford subalgebras, $\gamma^a$ and $\tilde{\gamma}^a$,
to the one described by $\gamma^a$~\cite{nh02,norma93,JMP2013}
%
\begin{eqnarray}
\{\tilde{\gamma}^a B &=&(-)^B\, i \, B \gamma^a\}\, |\psi_{oc}>\,,
\label{tildegammareduced0}
\end{eqnarray}
with $(-)^B = -1$, if $B$ is (a function of) odd products of $\gamma^a$'s, 
otherwise $(-)^B = 1$~\cite{nh02}, the vacuum state $|\psi_{oc}>$ will be 
defined in Eq.~(\ref{vaccliffodd}). 
%


The operators $\tilde{\gamma}^a$'s can after the postulate, Eq.~(\ref{tildegammareduced0}), be used to describe the quantum 
numbers of each of the $2^{\frac{d}{2}-1}$ irreducible representations (with
$2^{\frac{d}{2}-1}$ members each, representing ``families'' of fermions) 
of the Lorentz group in the internal spaces of fermions with the infinitesimal generators $S^{ab}$~\footnote{%
Lorentz group has in the internal space of fermions the generators $S^{ab}=$ 
$\frac{i}{4}(\gamma^a \gamma^b - \gamma^b \gamma^a)$)}, while  $\tilde{S}^{ab}$~\footnote{%
$\tilde{S}^{ab}$ $=\frac{i}{4}(\tilde{\gamma}^a 
\tilde{\gamma}^b - \tilde{\gamma}^b \tilde{\gamma}^a)$.}, 
transform a family member of one family to the same family 
member of all the families. 
We shall see~\cite{n2023NPB} that the quantum numbers of each irreducible representation of the Lorentz group in the internal space of bosons are equal to
${\cal S}^{ab}$ ($= S^{ab} + \tilde{S}^{ab}$)~\footnote{
One can prove (or read in App.~I of Ref.~\cite{nh2021RPPNP}) that the
relations of Eq.~(\ref{gammatildeantiher}) remain valid also after the
{\it postulate}, presented in Eq.~(\ref{tildegammareduced0}).}.

We shall arrange all the ``basis vectors'' describing internal spaces of fermion 
and boson second quantized fields to be the eigenstates of the Cartan 
subalgebra members of the Lorentz algebra, chosen to be
\begin{small}
\begin{eqnarray}
&&S^{03}, S^{12}, S^{56}, \cdots, S^{d-1 \;d}\,, \nonumber\\
&&\tilde{S}^{03}, \tilde{S}^{12}, \tilde{S}^{56}, \cdots,  \tilde{S}^{d-1\; d}\,, 
\nonumber\\
&&{\cal {\bf S}}^{ab} = S^{ab} +\tilde{S}^{ab}\,. 
\label{cartangrasscliff}
\end{eqnarray}
\end{small}
In even-dimensional spaces there are $\frac{d}{2}$ members of the Cartan
subalgebra, Eq.~(\ref{cartangrasscliff})~\footnote{In odd-dimensional spaces
there are $\frac{d-1}{2}$ members of the Cartan subalgebra.}.

We define~\cite{norma93,nh02} in even dimensional spaces $\frac{d}{2}$ 
nilpotents, $\stackrel{ab}{(k)}$, each nilpotent is a superposition of an
odd number of $\gamma^{a}$'s, and  projectors, $\stackrel{ab}{[k]}$, each
is a superposition of an even number of $\gamma^{a}$'s, 
\begin{small}
\begin{eqnarray}
\label{nilproj}
\stackrel{ab}{(k)}:&=&\frac{1}{2}(\gamma^a +
\frac{\eta^{aa}}{ik} \gamma^b)\,, \;\;\; (\stackrel{ab}{(k)})^2=0\, , 
\nonumber \\
\stackrel{ab}{[k]}:&=&
\frac{1}{2}(1+ \frac{i}{k} \gamma^a \gamma^b)\,, \;\;\;
(\stackrel{ab}{[k]})^2=\stackrel{ab}{[k]},
\end{eqnarray}
\end{small}
each nilpotent and projector is chosen to be in even-dimensional spaces the eigenvector of one of 
$\frac{d}{2}$ members of the Cartan subalgebra
\begin{small}
\begin{eqnarray}
\label{signature0}
S^{ab} \,\stackrel{ab}{(k)} = \frac{k}{2} \,\stackrel{ab}{(k)}\,,\quad && \quad
\tilde{S}^{ab}\,\stackrel{ab}{(k)} = \frac{k}{2} \,\stackrel{ab}{(k)}\,,\nonumber\\
S^{ab}\,\stackrel{ab}{[k]} = \frac{k}{2} \,\stackrel{ab}{[k]}\,,\quad && \quad
\tilde{S}^{ab} \,\stackrel{ab}{[k]} = - \frac{k}{2} \,\,\stackrel{ab}{[k]}\,,
\end{eqnarray}
\end{small}
with $k^2=\eta^{aa} \eta^{bb}$, pointing out that the eigenvalues of 
$S^{ab}$ on projectors expressed with $\gamma^a$ differ from the eigenvalues 
of $\tilde{S}^{ab}$ on projectors expressed with $\gamma^a$.
Taking into account Eq.~(\ref{gammatildeantiher}) one finds the relations
\begin{small}
\begin{eqnarray}
\label{usefulrel0}
\gamma^a \stackrel{ab}{(k)}&=& \eta^{aa}\stackrel{ab}{[-k]},\;\;
\gamma^b \stackrel{ab}{(k)}= -ik \stackrel{ab}{[-k]}, \; \; 
\gamma^a \stackrel{ab}{[k]}= \stackrel{ab}{(-k)},\; \;\;
\gamma^b \stackrel{ab}{[k]}= -ik \eta^{aa} \stackrel{ab}{(-k)}\,,\nonumber\\
\tilde{\gamma^a} \stackrel{ab}{(k)} &=& - i\eta^{aa}\stackrel{ab}{[k]},\;\;
\tilde{\gamma^b} \stackrel{ab}{(k)} = - k \stackrel{ab}{[k]}, \;\; \,
\tilde{\gamma^a} \stackrel{ab}{[k]} = \;\;i\stackrel{ab}{(k)},\; \quad
\tilde{\gamma^b} \stackrel{ab}{[k]} = -k \eta^{aa} \stackrel{ab}{(k)}\,,
\nonumber\\ %
\stackrel{ab}{(k)}\stackrel{ab}{(-k)}& =& \eta^{aa} \stackrel{ab}{[k]}\,,\quad \;
\stackrel{ab}{(-k)}\stackrel{ab}{(k)} = \eta^{aa} \stackrel{ab}{[-k]}\,,\quad\;
\stackrel{ab}{(k)}\stackrel{ab}{[k]} =0\,,\quad \quad\,
\stackrel{ab}{(k)}\stackrel{ab}{[-k]} =
\stackrel{ab}{(k)}\,,\quad 
\nonumber\\ 
\stackrel{ab}{(-k)}\stackrel{ab}{[k]} &=& \stackrel{ab}{(-k)}\,,\quad \quad\quad
\stackrel{ab}{[k]}\stackrel{ab}{(k)}= \stackrel{ab}{(k)}\,,
\quad \quad \quad \;
\stackrel{ab}{[k]}\stackrel{ab}{(-k)} =0\,,\quad \quad\,
\stackrel{ab}{[k]}\stackrel{ab}{[-k]} =0\,,\quad
\nonumber\\
\stackrel{ab}{(k)}^{\dagger} &=& \eta^{aa}\stackrel{ab}{(-k)}\,,\quad
(\stackrel{ab}{(k)})^2 =0\,, \quad \stackrel{ab}{(k)}\stackrel{ab}{(-k)}
=\eta^{aa}\stackrel{ab}{[k]}\,,\nonumber\\
\stackrel{ab}{[k]}^{\dagger} &=& \,\stackrel{ab}{[k]}\,, \quad \quad \quad \quad
(\stackrel{ab}{[k]})^2 = \stackrel{ab}{[k]}\,,
\quad \stackrel{ab}{[k]}\stackrel{ab}{[-k]}=0\,.
\end{eqnarray}
\end{small}
More relations are presented in App.~A of Ref.~\cite{nh2023dec}.

{\it We define}~\cite{norma93,nh02,n2023NPB} {\it in even dimensional spaces the ``basis vectors'' of fermion and boson second quantized fields as algebraic,} 
$*_A$, {\it products of nilpotents and projectors so
 that each product is an eigenvector of all} $\frac{d}{2}$ {\it Cartan subalgebra 
 members,} Eq.~(\ref{cartangrasscliff}). 
 
Fermion ``basis vectors'' must be products of an odd 
number of nilpotents, at least one, the rest are projectors; boson ``basis vectors'' 
must be products of an even number of nilpotents, the rest are projectors~\footnote{
An odd product of nilpotents anti-commute with another product of an odd number of different 
nilpotents, explaining the anti-commutation postulates for fermions. To the creation 
operators, which are tensor products of the ``basis vectors'' and the basis in ordinary 
space, determine anti-commutativity the ``basis vectors''. Correspondingly the 
even products of nilpotents explain the commutation postulates for boson fields.}.\\

We recognize:
Half of $2^d$ different products of $\gamma^a$'s are odd, and half of them are 
even. The  odd ``basis vectors'' appear in $2^{\frac{d}{2}-1}$ irreducible 
representations, we call them families, each family having $2^{\frac{d}{2}-1}$ 
members (obtainable from any other member by $S^{ab}$; the family member 
of any other family is obtainable by $\tilde{S}^{ab}$). 
Since the Hermitian conjugated partner of a nilpotent 
$\stackrel{ab}{(k)}^{\dagger}$ is $\eta^{aa}\stackrel{ab}{(-k)}$, it follows 
that the Hermitian conjugated partners of the odd ``basis vectors'' with 
an odd number of nilpotents must belong to a different group of 
$ 2^{\frac{d}{2}-1}$ members in $2^{\frac{d}{2}-1}$ families.  \\

The  even ``basis vectors'' with an even number of nilpotents have
their Hermitian conjugated partners within the same group; projectors are self
adjoint, $S^{ac}$ transforms $\stackrel{ab}{(k)} *_A \stackrel{cd}{(k')}$
into $\stackrel{ab}{[-k]} *_A \stackrel{cd}{[-k']}$, while $\tilde{S}^{ac}$
transforms $\stackrel{ab}{(k)} *_A \stackrel{cd}{(k')}$ into
$\stackrel{ab}{[k]} *_A $ $ \stackrel{cd}{[k']}$, what follows if taking into 
account Eq.~(\ref{usefulrel0}). Since the number of the  odd products 
of $\gamma^a$'s is equal to the number of the  even products of 
$\gamma^a$'s, there must be another group of the  even ``basis 
vectors'' with $2^{\frac{d}{2}-1}\times 2^{\frac{d}{2}-1}$ members.

We learn~\cite{n2023NPB,n2024NPB} that one group of the  even 
``basis vectors'' transforms the family members of the  odd 
``basis vectors'' among
themselves, while the second group of the  even ``basis vectors'' 
transform any member of a family to the same member of another families.

Let us clear up that the algebraic application, $*_{A}$, of the  even 
``basis vectors'', we name them either  
${}^{I}{\hat{\cal A}}^{m \dagger}_{f }$ if they are of the first kind
or ${}^{II}{\hat{\cal A}}^{m \dagger}_{f }$  if they are of the second kind,
on the odd ``basis vectors'', we name them 
$ \hat{b}^{m' \dagger}_{f `} $, transforms the odd ``basis vectors''
into the odd ``basis vectors'', meaning that while the  odd 
``basis vectors''  carry the half integer values of the Cartan subalgebra 
members eigenvalues, $\pm \frac{i}{2}$ or $\pm \frac{1}{2}$, carry the 
 even "basis vectors" the eigenvalues of the Cartan subalgebra 
members 
$(\pm i, 0)$ or $(\pm 1,0)$, in agreement with ${\cal S}^{ab}$
$= (S^{ab} + \tilde{S}^{ab})$~\cite{n2023NPB,n2024NPB}.



%
\subsection{``Basis vectors'' in  $d=2(2n+1)$-dimensional internal spaces 
from the point of view of $d=(3+1)$}
\label{bosons13+1and5+1}


This part overviews several papers with the same
topic~(\cite{nh2021RPPNP,n2023MDPI,n2023NPB,n2024NPB} and references 
therein).\\

\vspace{2mm}

{\bf i.} {\it The odd ``basis vectors''}\\

Let us start in $d=2(2n+1)$ with the ``basis vector'' $\hat{b}^{1 \dagger}_{1}$
which is the product of only nilpotents, all the rest members belonging to the $f=1$
family follow by the application of $S^{ab}$, presented on the left-hand side of 
Eq.~(\ref{allcartaneigenvec}).
Their Hermitian conjugated partners, $\hat{b}^{m} _{f} =$
$(\hat{b}^{m \dagger}_{f})^{\dagger}$ , are presented on the right-hand 
side~\footnote{%
The algebraic product mark, $*_{A}$, among nilpotents and projectors is skipped.}.
\begin{small}
\begin{eqnarray}
\label{allcartaneigenvec}
&& \hat{b}^{1 \dagger}_{1}=\stackrel{03}{(+i)}\stackrel{12}{(+)} \stackrel{56}{(+)}
\cdots \stackrel{d-1 \, d}{(+)}\,,\qquad \qquad \qquad \quad \quad
\hat{b}^{1}_{1}=\stackrel{03}{(-i)}\stackrel{12}{(-)}\cdots \stackrel{d-1 \, d}{(-)}\,,
\nonumber\\
&&\hat{b}^{2 \dagger}_{1} = \stackrel{03}{[-i]} \stackrel{12}{[-]}
\stackrel{56}{(+)} \cdots \stackrel{d-1 \, d}{(+)}\,,\qquad \qquad \qquad \qquad\;\;
\hat{b}^{2 }_{1} = \stackrel{03}{[-i]} \stackrel{12}{[-]}
\stackrel{56}{(-)} \cdots \stackrel{d-1 \, d}{(-)}\,,\nonumber\\
&& \cdots \qquad \qquad \qquad \qquad \qquad \qquad \qquad \qquad \qquad\;
\cdots \nonumber\\
&&\hat{b}^{2^{\frac{d}{2}-1} \dagger}_{1} = \stackrel{03}{(+i)} \stackrel{12}{[-]}
\stackrel{56}{[-]} \dots \stackrel{d-3\,d-2}{[-]}\;\stackrel{d-1\,d}{[-]}\,, \quad
\hat{b}^{2^{\frac{d}{2}-1}}_{1} = \stackrel{03}{(-i)} \stackrel{12}{[-]}
\stackrel{56}{[-]} \dots \stackrel{d-3\,d-2}{[-]}\;\stackrel{d-1\,d}{[-]}\,,
\nonumber\\
&& \cdots\,, \qquad \qquad \qquad \qquad \qquad \qquad \qquad \qquad \cdots\,.
\end{eqnarray}
\end{small}
All the members on the left hand side are orthogonal among themselves, and all 
the members of the right hand side are orthogonal among themselves, due to
 Eq.~(\ref{usefulrel0})~\footnote{%
\begin{eqnarray}
\hat{b}^{m \dagger}_f *_{A} \hat{b}^{m `\dagger }_{f `}&=& 0\,,
\quad \hat{b}^{m}_f *_{A} \hat{b}^{m `}_{f `}= 0\,, \quad \forall m,m',f,f `\,.
\label{orthogonalodd}
\end{eqnarray}
Any member of $2^{\frac{d}{2}-1}$ families follow by the application of 
$\tilde{S}^{ab}$.

Choosing the vacuum state equal to
\begin{eqnarray}
\label{vaccliffodd}
|\psi_{oc}>= \sum_{f=1}^{2^{\frac{d}{2}-1}}\,\hat{b}^{m}_{f}{}_{*_A}
\hat{b}^{m \dagger}_{f} \,|\,1\,>\,,
\end{eqnarray}
for one of the members $m$, which can be anyone of the odd irreducible
representations $f$
it follows that the odd ``basis vectors'' obey the relations
\begin{eqnarray}
\label{almostDirac}
\hat{b}^{m}_{f} {}_{*_{A}}|\psi_{oc}>&=& 0.\, |\psi_{oc}>\,,\nonumber\\
\hat{b}^{m \dagger}_{f}{}_{*_{A}}|\psi_{oc}>&=&  |\psi^m_{f}>\,,\nonumber\\
\{\hat{b}^{m}_{f}, \hat{b}^{m'}_{f `}\}_{*_{A}+}|\psi_{oc}>&=&
 0.\,|\psi_{oc}>\,, \nonumber\\
\{\hat{b}^{m \dagger}_{f}, \hat{b}^{m' \dagger}_{f  `}\}_{*_{A}+}|\psi_{oc}>
&=& 0. \,|\psi_{oc}>\,,\nonumber\\
\{\hat{b}^{m}_{f}, \hat{b}^{m' \dagger}_{f `}\}_{*_{A}+}|\psi_{oc}>
&=& \delta^{m m'} \,\delta_{f f `}|\psi_{oc}>\,.
\end{eqnarray}
 }. 
 The anti-commutation relations  among the ``basis vectors'' and their Hermitian 
conjugated partners fulfil the anti-commutation relations postulated by Dirac
for the second quantized fermion fields, Eq.~(\ref{almostDirac}).\\

It is easy to reproduce all the matrices postulated by Dirac for all $S^{ab}$,
although it is not needed; the application of any $S^{ab}$ on any of the 
members of any of the families is very simple.~\footnote{%
Let be pointed out that the Dirac's matrices were constructed for the massive 
fermions, correspondingly they do not anticommute~\cite{tdn2014}
.}
 \\
The application of 
$\tilde{S}^{ab}$, they do not appear among the Dirac operators, are equally
simple. The application of $\gamma^a$'s, as well as of $\tilde{\gamma}^a$'s,
on these states needs presentation of the even ``basis vectors'', since
they change the odd ``basis vectors'' into the even ``basis 
vectors''~\cite{n2023NPB,n2024NPB}.\\

\vspace{2mm}

{\bf ii.} {\it The even ``basis vectors''}\\

The  even ``basis vectors'' appear in two orthogonal groups, named   
${}^{I}\hat{\cal A}^{m \dagger}_{f}$ and  
${}^{II}\hat{\cal A}^{m \dagger}_{f}$. Each group has
$2^{\frac{d}{2}-1}\times $ $2^{\frac{d}{2}-1}$ members~\footnote{
The members of one group can not be reached by the members of another
group by either $S^{ab}$'s or $\tilde{S}^{ab}$'s or both.}.

The generators $S^{ab}$ and $\tilde{S}^{ab}$ generate from the starting 
``basis vector'' of each group all the $2^{\frac{d}{2}-1} \times$ 
$2^{\frac{d}{2}-1}$ members.
Each group contains the Hermitian conjugated partners within the same group;
$2^{\frac{d}{2}-1}$ members of each group are products of only (self-adjoint)
projectors, with all the eigenvalues of the Cartan subalgebra members, 
${\cal S}^{ab}=(S^{ab} +\tilde{S}^{ab})$,  presented in 
Eq.~(\ref{cartangrasscliff}), equal zero.

\begin{eqnarray}
\label{allcartaneigenvecevenI}
{}^I\hat{{\cal A}}^{1 \dagger}_{1}=\stackrel{03}{(+i)}\stackrel{12}{(+)}\cdots
\stackrel{d-1 \, d}{[+]}\,,\qquad &&
{}^{II}\hat{{\cal A}}^{1 \dagger}_{1}=\stackrel{03}{(-i)}\stackrel{12}{(+)}\cdots
\stackrel{d-1 \, d}{[+]}\,,\nonumber\\
{}^I\hat{{\cal A}}^{2 \dagger}_{1}=\stackrel{03}{[-i]}\stackrel{12}{[-]}
\stackrel{56}{(+)} \cdots \stackrel{d-1 \, d}{[+]}\,, \qquad &&
{}^{II}\hat{{\cal A}}^{2 \dagger}_{1}=\stackrel{03}{[+i]}\stackrel{12}{[-]}
\stackrel{56}{(+)} \cdots \stackrel{d-1 \, d}{[+]}\,,
\nonumber\\
{}^I\hat{{\cal A}}^{3 \dagger}_{1}=\stackrel{03}{(+i)} \stackrel{12}{(+)}
\stackrel{56}{(+)} \cdots \stackrel{d-3\,d-2}{[-]}\;\stackrel{d-1\,d}{(-)},&&
{}^{II}\hat{{\cal A}}^{3 \dagger}_{1}=\stackrel{03}{(-i)} \stackrel{12}{(+)}
\stackrel{56}{(+)} \cdots \stackrel{d-3\,d-2}{[-]}\;\stackrel{d-1\,d}{(-)}, \nonumber\\
\dots \qquad && \dots 
\end{eqnarray}
%

The even ``basis vectors'' belonging to two different groups are orthogonal due
to the fact that they differ in the sign of one nilpotent or one projector or the algebraic
product of a member of one group with a member of another group gives zero according
to the third and fourth lines of Eq.~(\ref{usefulrel0}): $\stackrel{ab}{(k)}\stackrel{ab}{[k]} =0$,
$\stackrel{ab}{[k]}\stackrel{ab}{(-k)} =0$,
$\stackrel{ab}{[k]}\stackrel{ab}{[-k]} =0$.
\begin{eqnarray}
\label{AIAIIorth}
{}^{I}{\hat{\cal A}}^{m \dagger}_{f} *_A {}^{II}{\hat{\cal A}}^{m \dagger}_{f}
&=&0={}^{II}{\hat{\cal A}}^{m \dagger}._{f} *_A
{}^{I}{\hat{\cal A}}^{m \dagger}_{f}\,.
\end{eqnarray}
The members of each of these two groups have the property.
\begin{eqnarray}
\label{ruleAAI}
{}^{i}{\hat{\cal A}}^{m \dagger}_{f} \,*_A\, {}^{i}{\hat{\cal A}}^{m' \dagger}_{f `}
\rightarrow \left \{ \begin{array} {r}
{}^{i}{\hat{\cal A}}^{m \dagger}_{f `}\,, i=(I,II) \\
{\rm or \,zero}\,.
\end{array} \right.
\end{eqnarray}
For a chosen ($m, f, f `$) there is only one $m'$ (out of $2^{\frac{d}{2}-1}$)
which gives a nonzero contribution.


Two ``basis vectors'', ${}^{i}{\hat{\cal A}}^{m \dagger}_{f}$ and
${}^{i}{\hat{\cal A}}^{m' \dagger}_{f '}$, the algebraic product, $*_{A}$, 
of which gives non zero contribution, ``scatter'' into the third one
${}^{i}{\hat{\cal A}}^{m \dagger}_{f `}$, for $i=(I,II)$.\\



\subsection{The relations among the odd and the even ``basis vectors''}
\label{relationsCliffordoddeven}

The algebraic application, $*_{A}$, of the even ``basis vectors''
${}^{I}{\hat{\cal A}}^{m \dagger}_{f }$ on the odd ``basis vectors''
$ \hat{b}^{m' \dagger}_{f `} $ and the  odd ``basis vectors''
$ \hat{b}^{m \dagger}_{f } $ on ${}^{II}{\hat{\cal A}}^{m \dagger}_{f }$
 gives%
\begin{eqnarray}
\label{calIAbbIIA1234gen}
{}^{I}{\hat{\cal A}}^{m \dagger}_{f } \,*_A \, \hat{b}^{m' \dagger }_{f `}
\rightarrow \left \{ \begin{array} {r} \hat{b }^{m \dagger}_{f `}\,, \\
{\rm or \,zero}\,,
\end{array} \right.\\
\hat{b}^{m \dagger }_{f } *_{A} {}^{II}{\hat{\cal A}}^{m' \dagger}_{f `} \,
\rightarrow \left \{ \begin{array} {r} \hat{b }^{m \dagger}_{f ``}\,, \\
{\rm or \,zero}\,,
\end{array} \right.
\end{eqnarray}
while
\begin{eqnarray}
\label{calbIA1234gen}
\hat{b}^{m \dagger }_{f } *_{A} {}^{I}{\hat{\cal A}}^{m' \dagger}_{f `} = 0
\,, \quad
{}^{II}{\hat{\cal A}}^{m \dagger}_{f } \,*_A \, \hat{b}^{m' \dagger }_{f `}= 0
\,,\;\;
\forall (m, m`, f, f `)\,.
\end{eqnarray}
%


Eq.~(\ref{calIAbbIIA1234gen}) demonstrates that
${}^{I}{\hat{\cal A}}^{m \dagger}_{f}$,
applying on $\hat{b}^{m' \dagger }_{f `} $, transforms the  odd
``basis vector'' into another odd ``basis vector'' of the same family,
transferring to the odd ``basis vector'' integer spins or gives zero.

Scattering of the odd ``basis vector'' $\hat{b}^{m \dagger }_{f }$ on
$ {}^{II}{\hat{\cal A}}^{m' \dagger}_{f `}$ transforms the 
odd
``basis vector'' into another odd ``basis vector''
$\hat{b }^{m \dagger}_{f ``}$ belonging to the same family member $m$
of a different family $f ``$.

\vspace{2mm}

{\it Both groups of even ``basis vectors'' manifest as the gauge fields
of the corresponding fermion fields: One concerning the family members
quantum numbers, the other concerning the family quantum numbers. }\\

\vspace{2mm}

The even ``basis vectors''  can be represented as algebraic products 
of the odd ``basis vectors'' and their Hermitian conjugated partners%
~\cite{n2024NPB}:

Knowing the odd ``basis vectors'' $\hat{b}^{m \dagger }_{f }$ of 
one family (any one) we are able to generate all the even 
$ {}^{I}{\hat{\cal A}}^{m' \dagger}_{f `}$ 
``basis vectors''
\begin{eqnarray}
\label{AIbbdagger}
{}^{I}{\hat{\cal A}}^{m \dagger}_{f}&=&\hat{b}^{m' \dagger}_{f `} *_A 
(\hat{b}^{m'' \dagger}_{f `})^{\dagger}\,.
\end{eqnarray}
Knowing the odd ``basis vectors'' $\hat{b}^{m \dagger }_{f }$ of 
one family member (any one) of all families we are able to generate all the even $ {}^{II}{\hat{\cal A}}^{m' \dagger}_{f `}$ 
``basis vectors'' 
\begin{eqnarray}
\label{AIIbdaggerb}
 {}^{II}{\hat{\cal A}}^{m \dagger}_{f}&=&
(\hat{b}^{m' \dagger}_{f `})^{\dagger} *_A 
\hat{b}^{m' \dagger}_{f `'}\,. 
\end{eqnarray}

\subsection{Relations among fermions and their vector and scalar gauge 
fields under the assumptions that all the gauge fields are active (have
non zero momentum) only in $d=(3+1)$}
\label{FermionsVetorScalar}

We will learn in Sect.~\ref{creationannihilation} that all the  even
``basis vectors'' have to carry the space index $\alpha$ which is equal to
$\mu=(0,1,2,3)$ if they describe the vector component of the ``basis
vectors'', and they are equal to $\sigma=(5,6,...)$ if describing the scalar
components of the ``basis vectors''.\\

Let us start with few examples. \\


{\bf a.}
The $d=(1+1)$ is very simple. We present it since we shall use it when 
trying to extend point fermion and boson fields into strings.

We have, in this case, two odd and two even eigenvectors of
the Cartan subalgebra members
\begin{small} 
 \begin{eqnarray}
 \label{1+1oddeven}
 && {\rm  \;\;odd}\nonumber\\
 \hat{b}^{ 1 \dagger}_{1}&=&\stackrel{01}{(+i)}\,, \quad 
 \hat{b}^{ 1 }_{1}=\stackrel{01}{(-i)}\,,\nonumber\\
 &&{\rm \;\;even} \;\nonumber\\
 {}^{I}{\bf {\cal A}}^{1 \dagger}_{1}&=&\stackrel{01}{[+i]}\,, \quad 
{}^{II}{\bf {\cal A}}^{1 \dagger}_{1}=\stackrel{01}{[-i]}\,.
 \end{eqnarray}
 \end{small}
The two odd ``basis vectors'' are Hermitian conjugated
to each other. The choice is made that  $\hat{b}^{ 1 \dagger}_{1}=
\stackrel{01}{(+i)}$ is the ``basis vector'', and the second 
odd object is its Hermitian conjugated partner. There is only one family
($2^{\frac{d}{2}-1}=1$) with one member. The vacuum state,
Eq.(\ref{vaccliffodd}), is for this choice equal to
$|\psi_{oc}>=\stackrel{01}{[-i]}$. The eigenvalue $S^{01}$ of
$\hat{b}^{ 1 \dagger}_{1}(= \stackrel{01}{(+i)})$ is
$\frac{i}{2}$.

Each of the two even ``basis vectors'' is self adjoint
($({}^{I,II}{\bf {\cal A}}^{1 \dagger}_{1})^{\dagger}=$
${}^{I,II}{\bf {\cal A}}^{1 \dagger}_{1}$), with the eigenvalues
${\cal S}^{01}$ equal  to $0$.\\
 We find, according to Eqs.~(\ref{AIbbdagger}, \ref{AIIbdaggerb}, \ref{usefulrel0}),
that
$$ {}^{I}{\bf {\cal A}}^{1 \dagger}_{1}=  \hat{b}^{ 1 \dagger}_{1}\,*_A\,
(\hat{b}^{ 1 \dagger}_{1})^{\dagger}\, ,\quad\,\quad
{}^{II}{\bf {\cal A}}^{1 \dagger}_{1} = (\hat{b}^{ 1 \dagger}_{1})^{\dagger}\,*_A\, \hat{b}^{ 1 \dagger}_{1}.$$


\vspace{2mm}

{\bf b.} The case with $d=2(2n+1)$, with $n=1$, is more illustrative.
App.~\ref{tables} presents the odd and even ``basis vectors'' in
Table~\ref{Table Clifffourplet.}.\\
We have in this case $16$ odd  ``basis vectors''; $4$ families with $4$
members each, and $16$ their Hermitian conjugated partners.

Each family can represent the internal spaces of ``positron'' with positive
``charge'', $S^{56}= \frac{1}{2}$, and of an ``electron'' with negative
``charge'', $S^{56}=- \frac{1}{2}$, as can be seen in
Table~\ref{Table Clifffourplet.}. The ``basis vectors'' of the ``positron''
and ``electron'' are
related by the charge conjugation operator~\footnote{%
In Ref.~\cite{nhds}, the discrete symmetry operators for fermion fields
in $d=2(2n +1)$ with the desired properties in $d=(3+1)$ are discussed.},
which includes in $d=(5+1)$ operators $\gamma^0 \gamma^5 $,
transforming $\hat{b}^{1 \dagger}_{f}$ into $\hat{b}^{1 \dagger}_{f}$,
of each family $f$.\\
The even ``basis vectors''  appear in Table~\ref{Table Clifffourplet.}
in two orthogonal groups, each group has $16$ members, their Hermitian
conjugated partners appear within the same group. The eigenvalues of the
Cartan subalgebra members ${\cal S}^{ab}=(S^{ab}+ \tilde{S}^{ab})$
are equal to either $(\pm i,0)$ or $(\pm 1,0)$.

In Tables~(2, 3, 4, 5) of Ref.~\cite{n2024NPB}, all the $32$ 
even ``basis vectors'' are expressed as the algebraic products of the odd ``basis vectors'' and their Hermitian conjugated partners. Two of these
tables, Tables~(\ref{S120Cliff basis5+1even I.},
\ref{transverseCliff basis5+1even I.}), are presented also in
App.~\ref{tables} of this contribution. 

We will learn in Sect.~\ref{creationannihilation} that all the  even
``basis vectors'' have to carry the space index $\alpha$ which is equal to
$\mu=(0,1,2,3)$ if they describe the vector component of the ``basis
vectors'', and they are equal to $\sigma=(5,6)$ if describing the scalar
components of the ``basis vectors''.\\

Let us illustrate these two groups of the even ``basis vectors'':\\

Sixteen ${}^{I}{\bf {\cal A}}^{m \dagger}_{f}$ transform family members
of any of $4$ families among themselves.
One can check that the even ``basis vectors'' can be written as
the algebraic products of the odd ``basis vectors'' and their
Hermitian conjugated  partners, Eqs.~(\ref{AIbbdagger}, \ref{AIIbdaggerb});
${}^{I}{\hat{\cal A}}^{m \dagger}_{f}=\hat{b}^{m' \dagger}_{f `} *_A
(\hat{b}^{m'' \dagger}_{f `})^{\dagger}$.

To describe ``photons'', the ``basis vectors''
$ {}^{I}{\hat{\cal A}}^{m \dagger}_{f}$ must not carry any non
zero eigenvalues of the Cartan
subalgebra members, Eq.~(\ref{cartangrasscliff}), since the ``photons''
can give to ``fermions'' (to ``electrons'' and ``positrons'' in our case)
only the spin offered by their space index $\mu$.
Table~\ref{S120Cliff basis5+1even I.} presents ``basis vectors'' for four
``photons'', marked in Table~\ref{S120Cliff basis5+1even I.} with
$\bigcirc$. All are selfadjoint operators, offering to ``basis vectors'' of 
``electrons'' and ``positrons'' no spin and no charge.

The remaining four even ``basis vectors'', appearing in the
Hermitian conjugated pairs (marked either by $\bigtriangleup$ or by
$\bullet$), would be allowed only if  ``fermions'' and ``bosons'' have
the non zero momenta in all six dimensions~\footnote{%
The even ``basis vectors'' including two nilpotents,
$\stackrel{03}{(\pm i)} \stackrel{56}{(\pm 1)}$, would transform 
``electrons'' into ``positrons'' and back, changing spin ${\cal S}^{03}$
and charge.}.
\\
To describe ``gravitons'', the ``basis vectors''
$ {}^{I}{\hat{\cal A}}^{m \dagger}_{f}$ must be able to offer
the internal spin ${\cal S}^{12}$, $(\pm1),$ (with ${\cal S}^{03},$
 $ (\pm i)$), together with  the spin offered by their space index 
$\mu$, while ${\cal S}^{56}$ eigenvalue must be equal to zero 
(must be described by a projector, not to be able to change the 
charge of ``electrons'' and ``positrons'').
Table~\ref{transverseCliff basis5+1even I.} presents ``basis 
vectors'' for four ``gravitons'', appearing in two Hermitian conjugated 
pairs, marked in Table~\ref{transverseCliff basis5+1even I.} either 
by $\ddagger$ or by $\odot \odot$.

The remaining four even ``basis vectors'', 
$ {}^{I}{\hat{\cal A}}^{m \dagger}_{f}$, appearing again in 
the Hermitian conjugated pairs (marked either by $\star \star$ or by
$\otimes$), would be allowed only if  ``fermions'' and ``bosons'' have
the non zero momenta in all six dimensions~\footnote{%
These even ``basis vectors'', they include two nilpotents
$\stackrel{12}{(\pm 1)} \stackrel{56}{(\pm 1)}$, would transform 
``electrons'' into ``positrons'' and back, changing the spin 
${\cal S}^{12}$ and charge.}.
\\

Sixteen ${}^{II}{\bf {\cal A}}^{m \dagger}_{f}$ transform a 
family member $m$ (any member) of a family $f$ to the same 
family member $m$ of all $4$ families.
One can check that the even ``basis vectors'' can be written 
as the algebraic products of the odd ``basis vectors'' and 
their Hermitian conjugated  partners, Eq.~(
\ref{AIIbdaggerb}); ${}^{II}{\hat{\cal A}}^{m \dagger}_{f}=
(\hat{b}^{m' \dagger}_{f `})^{\dagger} *_A
\hat{b}^{m'' \dagger}_{f `}$.

In Tables~(4,5) of Ref.~\cite{n2024NPB} all the sixteen members 
are presented. 

Also ${}^{II}{\bf {\cal A}}^{m \dagger}_{f}$ 
carry the space index $\alpha$ when taking part in creation 
operators. The space index $\alpha$ must be $\mu=(0,1,2,3)$ when
representing vector fields and $\sigma=(5,6)$ when representing 
scalars~\footnote{%
In Ref.~(\cite{nh2021RPPNP}, and reference therein) the role of 
the scalars in the realistic case $d=(13+1)$, suggested by 
experiments while representing Higgs's scalars, is explained.
}.\\

If fermions and bosons are active (have non zero momenta) only 
in $d=(3+1)$ ``gravitons'' and ``photons'' 
${}^{II}{\bf {\cal A}}^{m \dagger}_{f}$ are only possible. \\

\begin{small}
Let us present here one case among the $32$ cases, presented in 
Table~5 of Ref.~\cite{n2024NPB}.

$${}^{II}\hat{\cal A}^{4 \dagger}_1
 (\equiv \stackrel{03}{[-i]}\,\stackrel{12}{[-]} \stackrel{56}{[-]})=
(\hat{b}^{1 \dagger}_{4})^{\dagger} *_{A} \hat{b}^{1\dagger}_{4}\\
(\equiv \stackrel{03}{(-i)}\,\stackrel{12}{(-)} \stackrel{56}{(-)} *_{A} 
\stackrel{03}{(+i)}\,\stackrel{12}{(+)} \stackrel{56}{(+)}).$$
This can easily be checked if taking into account  
Eq.~(\ref{gammatildeantiher}) or Eq.~(\ref{usefulrel0}). 

One can check also that 
$ \hat{b}^{1\dagger}_{4}\,*_A\,{}^{II}\hat{\cal A}^{4 \dagger}_1=
\hat{b}^{1\dagger}_{4}$. The ``basis vector'' 
${}^{II}\hat{\cal A}^{4 \dagger}_1$ with $ {\cal S}^{03}=
{\cal S}^{12}={\cal S}^{56}=0$ can transfer to an ``electron'' or 
``positron'' only its space momentum while it cannot change their 
internal properties.
\end{small}
%

%
\subsection{The case with $d=2(2n+1)$, with $n=3$, is what the 
experiments suggest}
\label{d=(13+1)}

We have learned so far that to know all the 
$2^{\frac{d}{2}-1}\times 2^{\frac{d}{2}-1}$ members of the 
even ``basis vectors'' ${}^{I}\hat{\cal A}^{m \dagger}_f$ 
we need to know all the members of one (any one) family (one 
irreducible representation) of the odd ``basis vectors'' 
$ \hat{b}^{m\dagger}_{f}$, Eq.~(\ref{AIbbdagger}) (and their 
Hermitian conjugated partners). 

To know all the $2^{\frac{d}{2}-1}\times 2^{\frac{d}{2}-1}$ 
members of the even ``basis vectors'' 
${}^{II}\hat{\cal A}^{m \dagger}_f$ we need to know one 
family member (any one) of all the families, 
Eq.~(\ref{AIIbdaggerb}) (and their Hermitian conjugated 
partners). 

Before generating the even ``basis vectors'' as the 
algebraic products, $*_A$, of members of the odd 
``basis vectors'', representing quarks and leptons and 
antiquarks and antileptons, let us shortly overview the properties 
of one irreducible representation of quarks and leptons appearing 
together with the antiquark and antileptons in 
Table~\ref{Table so13+1.} of App.~\ref{13+1representation}.\\

In the even dimensional space $d=(13+1)$~(\cite{n2022epjc}, 
App.~A), one irreducible representation of the odd ``basis 
vectors'' if analysed from the point of view of the subgroups of the 
group $SO(13,1)$ (including $SO(7,1)\times SO(6)$, while $SO(7,1)$
breaks into $SO(3,1) \times SO(4)$, and $SO(6)$ breaks into 
$SU(3)\times U(1)$) contains the odd ``basis vectors'' 
describing internal spaces of quarks and leptons and antiquarks and 
antileptons, manifesting at low energies the quantum numbers 
assumed by the {\it standard model} before the electroweak break. 
Since $SO(4)$ contains two $SU(2)$ subgroups, $SU(2)_I$ and 
$SU(2)_{II}$, with the hypercharge of the {\it standard model} 
$Y=\tau^{23} + \tau^4$ (with $\tau^{23}$ belonging to 
$SU(2)_{II}$ and $\tau^4$ originating in $SO(6)$, breaking to 
$SU(3)\times U(1)$), 
one irreducible representation includes the right-handed neutrinos 
and the left-handed antineutrinos, which are not in the {\it standard 
model} scheme~\footnote{%
For an overview of the properties of the vector and scalar gauge 
fields in the {\it spin-charge-family} theory, the reader is invited to 
reed Refs.~(\cite{nh2021RPPNP,nd2017,n2023NPB,gmdn2007,%
gn2013} and the references therein). The reader can find in 
Table~\ref{Table so13+1.} that ``basis vectors'' of quarks have 
identical content of $SO(7,1)$ as ``basis vectors'' of leptons (and 
antiquarks as antiletons); they differ only in the 
$SU(3)\times U(1)$.}.\\ 

In Table~\ref{Table so13+1.}, one can read the quantum numbers 
of the odd ``basis vectors'' representing quarks and 
leptons {\it and antiquarks and antileptons} if taking into account 
that all the nilpotents and projectors are eigenvectors of one of 
the Cartan subalgebra members, ($S^{03}, S^{12}, S^{56}, 
\dots, $$S^{13\,14}$), with the eigenvalues $\pm \frac{i}{2}$ 
or $\pm \frac{1}{2}$, while for (${\cal S}^{03}, {\cal S}^{12},
{\cal S}^{56}, \cdots, {\cal S}^{13\,14}$) the eigenvalues
for $\stackrel{ab}{(\pm i)}$ are $\, \pm i$,  and for 
$\stackrel{ab}{(\pm )}\,$ are $ \pm 1$, and for 
$\stackrel{ab}{[\pm i]}$ and $\stackrel{ab}{[\pm ]}$ the 
eigenvalues are zero~\footnote{%
Let us remind the reader the difference between eigenvectors of
$S^{ab}$ and $\tilde{S}^{ab}$: Applying on a nilpotent they
both give the same eigenvalue, while they give opposite 
eigenvalue if applying on a projector, Eq.~(\ref{signature0}).
}.

Taking into account that the third component of the {\it standard 
model} weak charge, $\tau^{13}=\frac{1}{2} (S^{56}-S^{78})$, 
of the third component of the second $SU(2)_{II}$ charge 
(not appearing in the {\it standard model}) $\tau^{23}=
\frac{1}{2} (S^{56}+ S^{78})$, of the colour charge 
[$\tau^{33}=\frac{1}{2} (S^{9\, 10}-S^{11\,12})$ and 
$\tau^{38}=\frac{1}{2\sqrt{3}} (S^{9\, 10}+S^{11\,12} - 
2 S^{13\,14})$], of the ``fermion charge'' 
$\tau^4=-\frac{1}{3} (S^{9\, 10}+S^{11\,12} + S^{13\,14})$, 
of the hyper charge $Y=\tau^{23} + \tau^4$, and of the 
electromagnetic charge $Q=Y + \tau^{13}$, one reproduces 
all the quantum  numbers of quarks, leptons, and  antiquarks, and 
antileptons. \\

\vspace{2mm}


Let us represent internal spaces (that is the ``basis vectors'') of 
photons, gravitons, weak bosons and coloured bosons in the way 
we learned in the cases {\,\bf b.} of Subsect.~\ref{FermionsVetorScalar}, that is as algebraic products of 
the odd ``basis vectors'' (represented in 
Table~\ref{Table so13+1.}  as $u_{R}^{ci}, d_{R}^{ci},  
u_{L}^{ci},  d_{L}^{ci} $ for quarks of colour $ci$, as
$\bar{u}_{L}^{\bar{ci}}, \bar{d}_{L}^{\bar{ci}}, 
\bar{u}_{R}^{\bar{ci}}, \bar{d}_{R}^{\bar{ci}}$ for antiquarks
of the opposite colour, as $\nu_{R}, e_{R}, \nu_{L}, e_{L}$ for 
leptons, $\bar{\nu}_{L}, \bar{e}_{L}, \bar{\nu}_{R}, \bar{e}_{R}$  
for the corresponding antileptons, their Hermitian conjugated 
partners will be written as $(\:\,)^{\dagger}$).\\


Starting from photons we need to have in mind that photons can 
give to quarks and leptons only momentum in ordinary space,
but cannot change internal spaces of quarks and leptons.

Let us look in Table~\ref{Table so13+1.} for anti-quark of
$u^{c1}_{L}$, seventh 
 line~(\cite{n2024NPB}, Eq.~(46)). The photon  
 ${}^{I}{\hat{\cal A}}^{ \dagger}_{ph \,\bar{u}^{\bar{c1}}_{R}
 \rightarrow\bar{u}^{\bar{c1}}_{R}}$ interacts with 
 $\bar{u}^{\bar{c1}}_{R\, 39^{th}}$ as  follows
\begin{small}
\begin{eqnarray}
\label{phAqeqe}
{}^{I}{\hat{\cal A}}^{ \dagger}_{ph \,\bar{u}^{\bar{c1} }_{R}
 \rightarrow \bar{u}^{\bar{c1}}_{R}}
(\equiv \stackrel{03}{[+i]}\stackrel{12}{[+]} \stackrel{56}{[-]}
 \stackrel{7 8}{[+]}\stackrel{9\,10}{[-]} \stackrel{11\,12}{[+]}
\stackrel{13\,14}{[+]})\,*_A\, \bar{u}^{\bar{c1}}_{R\, 39^{th}}
(\equiv \stackrel{03}{(+i)}\stackrel{12}{[+]} 
\stackrel{56}{(-)}\stackrel{78}{(+)}\stackrel{9\,10}{[-]} 
\stackrel{11\,12}{(+)} \stackrel{13\,14}{(+)})\,&&\nonumber\\
\rightarrow \bar{u}^{\bar{c1}}_{R\, 39^{th}}
(\equiv \stackrel{03}{(+i)}\stackrel{12}{[+]} 
\stackrel{56}{(-)}\stackrel{78}{(+)}\stackrel{9\,10}{[-]} 
\stackrel{11\,12}{(+)} \stackrel{13\,14}{(+)})\,, \; 
 {}^{I}{\hat{\cal A}}^{ \dagger}_{ph\, \bar{u}^{\bar{c1}}_{R}
 \rightarrow \bar{u}^{\bar{c1}}_{R}}= 
 {u}^{\bar{c1}}_{R\, 39^{th}}\,*_A\,
 ({u}^{\bar{c1}}_{R\, 39^{th}})^{\dagger}.&&\nonumber\\
%
\end{eqnarray}
\end{small}
Similarly, one finds photons interacting with the rest of  quarks and 
electrons and antiquarks and positrons~\footnote{%
The break of symmetry, discussed in Ref.~\cite{nh2021RPPNP}, but 
not yet really understood in this new way of presenting the internal
space of boson fields, prevents neutrinos to interact with photons.
Let us add that photons can not directly interact with another 
photons since the algebraic products of different selfadjoint 
operators are zero, but can interact with other boson fields, for 
example, with gravitons~\cite{n2024NPB}. However, algebraic 
product of the ``basis vector'' which is a product of projectors
by itself is the same ``basis vector'' back.}. \\


Having in mind that bosons have an even number of nilpotents,
and taking $u^{c1}_{R}$ from the first  and the second line of 
Table~\ref{Table so13+1.}, we find that the ``basis vector'' of 
the ``graviton'' 
${}^{I}{\hat{\cal A}}^{ \dagger}_{gr\, u^{c1}_{R\uparrow}
\rightarrow u^{c1}_{R\downarrow}}$ applies on 
$u^{c1}_{R\,1^{st}}$
as follows
\begin{small}
\begin{eqnarray}
\label{grAqeqe}
{}^{I}{\hat{\cal A}}^{ \dagger}_{gr\, u^{c1}_{R\uparrow}
\rightarrow u^{c1}_{R\downarrow}}
(\equiv \stackrel{03}{(-i)}\stackrel{12}{(-)} \stackrel{56}{[+]}
 \stackrel{7 8}{[+]}\stackrel{9\,10}{[+]} \stackrel{11\,12}{[-]}
\stackrel{13\,14}{[-]})\,*_A\, u^{c1}_{R\, 1^{st}}
(\equiv \stackrel{03}{(+i)}\stackrel{12}{[+]} 
\stackrel{56}{[+]}\stackrel{78}{(+)}\stackrel{9\,10}{(+)} 
\stackrel{11\,12}{[-]} \stackrel{13\,14}{[-]})\,&&
\nonumber\\
\rightarrow u^{c1}_{R\, 2^{nd}}
(\equiv \stackrel{03}{[-i]}\stackrel{12}{(-)} 
\stackrel{56}{[+]}\stackrel{78}{(+)}\stackrel{9\,10}{(+)} 
\stackrel{11\,12}{[-]} \stackrel{13\,14}{[-]})\,,
{}^{I}{\hat{\cal A}}^{ \dagger}_{gr\, u^{c1}_{R\uparrow}
\rightarrow u^{c1}_{R\downarrow}} = u^{c1}_{R\, 2^{nd}}
\,*_A\, (u^{c1}_{R\, 1^{st}})^{\dagger}\,,&&\nonumber\\
\end{eqnarray}
\end{small}
Similarly we can find the ``basis vectors'' of all gravitons 
transforming the  spins $S^{03}$ and $S^{12}$ in internal 
spaces of quarks and leptons. The gravitons carry in addition 
the momentum in ordinary space time.\\



Looking  for the weak bosons, which transform 
$d^{c1}_{L\, 5 {th}}$  (presented in 
Table~\ref{Table so13+1.} in $5^{th}$ line) to  
$u^{c1}_{L\,7^{th}}$ (presented in 
Table~\ref{Table so13+1.} in $7^{th}$ line), 
we find
\begin{small}
\begin{eqnarray}
\label{w1enuL}
{}^{I}{\hat{\cal A}}^{ \dagger}_{w1 \,d^{c1} _{L}
\rightarrow u^{c1}_{L}}
(\equiv \stackrel{03}{[-i]}\stackrel{12}{[+]} \stackrel{56}{(+)}
 \stackrel{7 8}{(-)}\stackrel{9\,10}{[+]} \stackrel{11\,12}{[-]}
\stackrel{13\,14}{[-]})\,*_A\, d^{c1}_{L\, 5 {th}}
(\equiv \stackrel{03}{[-i]}\stackrel{12}{[+]} 
\stackrel{56}{(-)}\stackrel{78}{(+)}\stackrel{9\,10}{(+)} 
\stackrel{11\,12}{[-]} \stackrel{13\,14}{[-]})\,&&
\nonumber\\
\rightarrow u^{c1}_{L\, 7 {th}}
(\equiv \stackrel{03}{[-i]}\stackrel{12}{[+]} 
\stackrel{56}{[+]}\stackrel{78}{[-]}\stackrel{9\,10}{(+)} 
\stackrel{11\,12}{[-]} \stackrel{13\,14}{[-]})
(\equiv \stackrel{03}{[-i]}\stackrel{12}{[+]} \,, 
{}^{I}{\hat{\cal A}}^{ \dagger}_{w1 \,d^{c1} _{L}
\rightarrow u^{c1}_{L}}= u^{c1}_{L\, 7 {th}}\, *_A\,
(d^{c1}_{L\, 5 {th}})^{\dagger}\,.&&\nonumber\\
\end{eqnarray}
\end{small}
We can find equivalently  the weak bosons ``basis vectors'' 
which cause transformations among other weak pairs.\\


Let us look  for the `basis vectors'' of gluons, transforming colour 
charges of quarks, for example, of $d^{c1}_{L\,5^{th}}$ 
(presented in Table~\ref{Table so13+1.} in $5^{th}$ line) to  
$d^{c3}_{L\,21^{st}}$ (presented in 
Table~\ref{Table so13+1.} in $21^{st}$ line)
%
\begin{small}
\begin{eqnarray}
\label{guc1c2R}
{}^{I}{\hat{\cal A}}^{ \dagger}_{gl\, d^{c1}_{L}
\rightarrow d^{c3}_{L}}
(\equiv \stackrel{03}{[-i]}\stackrel{12}{[+]} \stackrel{56}{[-]}
 \stackrel{7 8}{[+]}\stackrel{9\,10}{(-)} \stackrel{11\,12}{[-]}
\stackrel{13\,14}{(+)})\,*_A\, d^{c1}_{L\,5^{th}}
(\equiv \stackrel{03}{[-i]}\stackrel{12}{[+]} 
\stackrel{56}{(-)}\stackrel{78}{(+)}\stackrel{9\,10}{(+)} 
\stackrel{11\,12}{[-]} \stackrel{13\,14}{[-]})\,&& 
\nonumber\\
\rightarrow d^{c3}_{L\,21^{st}}
(\equiv \stackrel{03}{[-i]}\stackrel{12}{[+]} 
\stackrel{56}{(-)}\stackrel{78}{(+)}\stackrel{9\,10}{[-]} 
\stackrel{11\,12}{[-]} \stackrel{13\,14}{(+)})\,, 
\quad 
{}^{I}{\hat{\cal A}}^{ \dagger}_{gl\, d^{c1}_{L}
\rightarrow d^{c3}_{L}}= d^{c3}_{L\,21^{st}}\, *_A\,
( d^{c1}_{L\,5^{th}})^{\dagger}\,.&&\nonumber\\
\end{eqnarray}
\end{small}\\

Let us conclude this section by repeating what we have 
learned in it: This section presents the internal degrees 
of freedom as the odd ``basis vectors'' for fermion 
fields and as the even ``basis vectors'' for boson 
fields for internal spaces $d=2(2n+1), n=0,1$ and for 
$n=3$. 
The creation operators for both fields are tensor products
 of ``basis vectors'' and basis in ordinary space-time. 
The even ``basis vectors'' of boson fields carry 
in addition the space index $\alpha$, 
Sect.~\ref{creationannihilation}.

There is the same number of the ``basis vectors'' of 
fermion and boson fields, manifesting  a kind of supersymmetry.
While fermion fields appear in families and have their
Hermitian conjugated partners in a separate groups, 
appear boson ``basis vectors'' in two orthogonal groups,
with their Hermitian conjugated partners within each
group. 

The break of symmetries is needed, which makes the 
number of observed families of fermions and their 
observed gauge fields reduced~\cite{nh2021RPPNP}. 
But the assumption that the fermion and boson fields 
have non zero momenta only in $d=(3+1)$  
breaks the Lorentz symmetry; the rotations $M^{ms}=
L^{ms}+ S^{ms}$, $m=(0,1,2,3, s=(5,6,...)$, 
($ S^{ms}$ can be replaced by $ \tilde{S}^{ms}$ 
and ${\cal S}^{ms}$), for example, can not go.\\

To extend point fields to strings, we can use the ``basis 
vectors'' for the extended part, presented in 
Eq.~(\ref{1+1oddeven}). For the extension
of the fermion fields we can have
$ \hat{b}^{ 1 \dagger}_{1}=
\stackrel{01}{(+i)}$ and  for its Hermitian conjugated 
partner $\hat{b}^{ 1 }_{1}=
(\hat{b}^{ 1 \dagger}_{1})^{\dagger}=
\stackrel{01}{(-i)}$. For the two boson fields we have   
${}^{I}{\bf {\cal A}}^{1 \dagger}_{1}=
\stackrel{01}{[+i]}$ and 
${}^{II}{\bf {\cal A}}^{1 \dagger}_{1}=
\stackrel{01}{[-i]}$~\footnote{%
Let us repeat that knowing the fermion ``basis 
vectors'' allows us to find both kinds of ``boson fields'',
${}^{I}{\bf {\cal A}}^{1 \dagger}_{1}= 
\hat{b}^{ 1 \dagger}_{1}\,*_A\, 
(\hat{b}^{ 1 \dagger}_{1})^{\dagger}\, =
\stackrel{01}{(+i)}\,*_A\,\stackrel{01}{(-i)}=\stackrel{01}{[+i]}$, and
${}^{II}{\bf {\cal A}}^{1 \dagger}_{1} = 
(\hat{b}^{ 1 \dagger}_{1})^{\dagger}\,*_A\, 
\hat{b}^{ 1 \dagger}_{1}=\stackrel{01}{(-i)}\,*_A\,
\stackrel{01}{(+i)}=\stackrel{01}{[-i]}$.}.  
\\

Sect.~\ref{string} discusses a possible extension of point
fields to strings.

%
\section{Creation and annihilation operators for fermion and boson fields 
in $d=(3+1)$}
\label{creationannihilation}

To define the creation operators for fermion or boson fields, besides the 
``basis vectors'' defining the internal spaces of fermions and bosons,
the basis in ordinary space in momentum or coordinate representation is 
needed.
Let us shortly overview Ref.~\cite{n2024NPB}, Sect.~3. (The extended 
version is presented in Ref.~\cite{nh2021RPPNP} in Subsect.~3.3 and 
App. J.)

The momentum basis is continuously infinite
\begin{eqnarray}
\label{creatorp}
|\vec{p}>&=& \hat{b}^{\dagger}_{\vec{p}} \,|\,0_{p}\,>\,,\quad
<\vec{p}\,| = <\,0_{p}\,|\,\hat{b}_{\vec{p}}\,, \nonumber\\
<\vec{p}\,|\,\vec{p}'>&=&\delta(\vec{p}-\vec{p}')=
<\,0_{p}\,|\hat{b}_{\vec{p}}\; \hat{b}^{\dagger}_{\vec{p}'} |\,0_{p}\,>\,,
\nonumber\\
&&{\rm pointing \;out\;} \nonumber\\
<\,0_{p}\,| \hat{b}_{\vec{p'}}\, \hat{b}^{\dagger}_{\vec{p}}\,|\,0_{p}\, > &=&\delta(\vec{p'}-\vec{p})\,,
\end{eqnarray}
with the normalization $<\,0_{p}\, |\,0_{p}\,>=1$.
While the quantized operators $\hat{\vec{p}}$ and $\hat{\vec{x}}$ commute
$\{\hat{p}^i\,, \hat{p}^j \}_{-}=0$, $\{\hat{x}^k\,, \hat{x}^l \}_{-}=0$,
it follows for $\{\hat{p}^i\,, \hat{x}^j \}_{-}=i \eta^{ij}$. One 
correspondingly finds
\begin{small}
\begin{eqnarray}
\label{eigenvalue10}
<\vec{p}\,| \,\vec{x}>=<0_{\vec{p}}\,|\,\hat{b}_{\vec{p}}\;
\hat{b}^{\dagger}_{\vec{x}}
|0_{\vec{x}}\,>=(<0_{\vec{x}}\,|\,\hat{b}_{\vec{x}}\;
\hat{b}^{\dagger}_{\vec{p}} \,\,
|0_{\vec{p}}\,>)^{\dagger}\, &&\nonumber\\
<0_{\vec{p}}\,|\{\hat{b}^{\dagger}_{\vec{p}}\,, \,
\hat{b}^{\dagger}_{\vec{p}\,'}\}_{-}|0_{\vec{p}}\,>=0\,,
<0_{\vec{p}}\,|\{\hat{b}_{\vec{p}}, \,\hat{b}_{\vec{p}\,'}\}_{-}|0_{\vec{p}}\,>=0\,,
<0_{\vec{p}}\,|\{\hat{b}_{\vec{p}}, \,\hat{b}^{\dagger}_{\vec{p}\,'}\}_{-}|0_{\vec{p}}\,>=0\,,&&
\nonumber\\
<0_{\vec{x}}\,|\{\hat{b}^{\dagger}_{\vec{x}}, \,\hat{b}^{\dagger}_{\vec{x}\,'}\}_{-}|0_{\vec{x}}\,>=0\,,
<0_{\vec{x}}\,|\{\hat{b}_{\vec{x}}, \,\hat{b}_{\vec{x}\,'}\}_{-}|0_{\vec{x}}\,>=0\,,
<0_{\vec{x}}\,|\{\hat{b}_{\vec{x}}, \,\hat{b}^{\dagger}_{\vec{x}\,'}\}_{-}|0_{\vec{x}}\,>=0\,,&&
\nonumber\\
<0_{\vec{p}}\,|\{\hat{b}_{\vec{p}}, \,\hat{b}^{\dagger}_{\vec{x}}\}_{-}|0_{\vec{x}}\,>=
e^{i \vec{p} \cdot \vec{x}} \frac{1}{\sqrt{(2 \pi)^{d-1}}}\,,
<0_{\vec{x}}\,|\{\hat{b}_{\vec{x}}, \,\hat{b}^{\dagger}_{\vec{p}}\}_{-}|0_{\vec{p}}\,>=
e^{-i \vec{p} \cdot \vec{x}} \frac{1}{\sqrt{(2 \pi)^{d-1}}}\,.&&\nonumber\\
\end{eqnarray}
\end{small}
The creation operators for either fermion or boson fields must be tensor products,
$*_{T}$, of both contributions, the ``basis vectors'' describing the 
internal space of fermions or bosons and the basis in ordinary 
momentum or coordinate space.

The creation operators for a free massless fermion field of the energy
$p^0 =|\vec{p}|$, belonging to a family $f$ and to a superposition 
of family members $m$ applying on the extended vacuum state 
including both spaces, $|\psi_{oc}>\,*_{T}\, |0_{\vec{p}}>$,  
can be written as
\begin{small}
\begin{eqnarray}
\label{wholespacefermions}
{\bf \hat{b}}^{s \dagger}_{f} (\vec{p}) \,&=& \,
\sum_{m} c^{sm}{}_f (\vec{p}) \,\hat{b}^{\dagger}_{\vec{p}}\,*_{T}\,
\hat{b}^{m \dagger}_{f} \, \,.
\end{eqnarray}
\end{small}
The creation operators $ \hat{\bf b}^{s\dagger}_{f }(\vec{p}) $ and their
Hermitian conjugated partners annihilation operators
$\hat{\bf b}^{s}_{f }(\vec{p}) $, creating and annihilating the single 
fermion states, respectively, fulfil when applying the vacuum state,
$|\psi_{oc}> *_{T} |0_{\vec{p}}>$, the anti-commutation relations for 
the second quantized fermions, postulated by Dirac (Ref.~\cite{n2023NPB},
Sect.3), explaining the Dirac's second quantization postulates for fermions.
\begin{small}
\begin{eqnarray}
<0_{\vec{p}}\,|
\{ \hat{\bf b}^{s' }_{f `}(\vec{p'})\,,\,
\hat{\bf b}^{s \dagger}_{f }(\vec{p}) \}_{+} \,|\psi_{oc}> |0_{\vec{p}}>&=&
\delta^{s s'} \delta_{f f'}\,\delta(\vec{p}' - \vec{p})\,\cdot |\psi_{oc}>
\,,\nonumber\\
\{ \hat{\bf b}^{s' }_{f `}(\vec{p'})\,,\,
\hat{\bf b}^{s}_{f }(\vec{p}) \}_{+} \,|\psi_{oc}> |0_{\vec{p}}>&=&0\, \cdot \,
|\psi_{oc}> |0_{\vec{p}}>
\,,\nonumber\\
\{ \hat{\bf b}^{s' \dagger}_{f '}(\vec{p'})\,,\,
\hat{\bf b}^{s \dagger}_{f }(\vec{p}) \}_{+}\, |\psi_{oc}> |0_{\vec{p}}>&=&0\, \cdot
\,|\psi_{oc}> |0_{\vec{p}}>
\,,\nonumber\\
\hat{\bf b}^{s \dagger}_{f }(\vec{p}) \,|\psi_{oc}> |0_{\vec{p}}>&=&
|\psi^{s}_{f}(\vec{p})>\,,\nonumber\\
\hat{\bf b}^{s}_{f }(\vec{p}) \, |\psi_{oc}> |0_{\vec{p}}>&=&0\, \cdot\,
\,|\psi_{oc}> |0_{\vec{p}}>\,, \nonumber\\
|p^0| &=&|\vec{p}|\,.
\label{Weylpp'comrel}
\end{eqnarray}
\end{small}

The creation operators for boson gauge fields must carry the space index
$\alpha$, describing the $\alpha$ component of the boson field in the
ordinary space~ \cite{n2024NPB}, Eq.~(24)).
We, therefore, add the space index $\alpha$ as follows
\begin{eqnarray}
\label{wholespacebosons}
{\bf {}^{i}{\hat{\cal A}}^{m \dagger}_{f \alpha}} (\vec{p}) \,&=&
{}^{i}{\hat{\cal C}}^{ m}{}_{f \alpha} (\vec{p})\,*_{T}\,
{}^{i}{\hat{\cal A}}^{m \dagger}_{f} \, \,, i=(I,II)\,,
\end{eqnarray}
with ${}^{i}{\hat{\cal C}}^{ m}{}_{f \alpha} (\vec{p})=
{}^{i}{\cal C}^{ m}{}_{f \alpha}\,\hat{b}^{\dagger}_{\vec{p}}$, with
$\hat{b}^{\dagger}_{\vec{p}}$ defined in Eqs.~(\ref{creatorp}, \ref{eigenvalue10}).
We treat free massless bosons of momentum $\vec{p}$ and energy $p^0=|\vec{p}|$
and of particular ``basis vectors'' ${}^{i}{\hat{\cal A}}^{m \dagger}_{f}$'s which are
eigenvectors of all the Cartan subalgebra members.
The creation operators for boson gauge fields commute.
%

%
\section{Points in ordinary space-time extended to strings; 
representations of fermions and bosons in odd dimensional spaces}
\label{string}
One possibility to achieve renormalizability of the proposed 
theory might be, learning from string 
theories~\cite{Blumenhagen,Kevin},
by extending the points in ordinary space-time 
to strings.
A second possibility might be to find out what can offer the
internal odd-dimensional spaces, manifesting two groups 
of anti-commuting ``basis vectors'' and two groups of
commuting ``basis vectors'', as mentioned in
Sect.~\ref{introduction} and explained in
Ref.~\cite{n2023MDPI,FadeevPopov}. One of the two 
groups of either anti-commuting or commuting ``basis 
vectors'' manifest a kind of Fadeev-Popov ghost.

\vspace{2mm} 

Let us start with the first possibility: extending the points in
ordinary space-time to strings.

Let us try to extend the fermion and boson fields,
expressed as tensor products, $\,*_T\,$, of  the ``basis
vectors'' describing the internal spaces of fermion and
boson fields and the basis in ordinary space-time to
strings, in the hope of achieving renormalizability for the
theory of free massless fermion and boson fields. As we
discussed in Sect.~\ref{basisvectors0}, the number of
``basis vectors'' which represent internal spaces of
fermions and the number of their Hermitian conjugated
partners (both together have 
$2 \times 2^{\frac{d}{2}-1}\times$
$2^{\frac{d}{2}-1} $ members), equals to the number of ``basis
vectors'' representing the internal spaces of bosons,
appearing in two orthogonal groups,
$ 2^{d-1}$$\;$~\footnote{%
One group of boson ``basis vectors'', while applying to any
of members of fermion families, forms a member of
the same family. The second group of boson ``basis
vectors'', when applying to a particular member of a family
transforms this member into the same member of another
family, Subsect.~\ref{d=(13+1)}.}, 
manifesting a kind of supersymmetry. (This is, however, 
not the usual kind of supersymmetry requiring that each 
fermion has a partner with the same charge among bosons, 
and vice versa.


We shall see that a trial to extend point particles in ordinary
space-time to strings in a ``stringy way'' requires to introduce
the strings time, besides the ordinary space-time and
correspondingly also the corresponding ``basis vectors'' in
$d=(1+1)$ in a string.

We assume that fermion and boson fields are active (having
non-zero momenta) only in $d=(3+1)$~\footnote{%
This assumption is not needed, but it seems meaningful.}, 
while we choose $d=2(2n+1)$ in the internal space.  
The observations suggest $n=3$.

Let us first assume a simple starting action, simplifying the 
action in Ref~(\cite{nh2021RPPNP} and in the references 
therein) for free massless fermion fields, and the 
corresponding free massless boson fields in 
$d=2(2n+1)$-dimensional space, having non zero
momenta only in $d=(3+1)$, while taking into account the 
creation operators
for fermion and boson fields expressed by the corresponding
``basic vectors'', Eqs.~(\ref{wholespacefermions}, \ref{wholespacebosons})~\footnote{%
The fermion states $\psi$  are defined by
$\hat{\bf b}^{s \dagger}_{f }(\vec{x},x^0)=
\sum_{m} \,\hat{b}^{ m \dagger}_{f} \,  \int_{- \infty}^{+ \infty} \,
\frac{d^{d-1}p}{(\sqrt{2 \pi})^{d-1}} \, c^{m s }{}_{f}\;
(\vec{p}) \;  \hat{b}^{\dagger}_{\vec{p}}\;
e^{-i (p^0 x^0- \varepsilon \vec{p}\cdot \vec{x})}$, applying on the
vacuum state $|\psi_{oc}>\,*_{T}\, |0_{\vec{p}}>$.
%
The boson states are defined by
${\bf {}^{I}{\hat{\cal A}}^{m \dagger}_{f \mu}}
(\vec{x}, x^0)=\int_{- \infty}^{+ \infty} \,
\frac{d^{d-1}p}{(\sqrt{2 \pi})^{d-1}} \,
{}^{I}{\hat{\cal A}}^{m \dagger}_{f \mu}  (\vec{p})\,
e^{-i (p^0 x^0- \varepsilon \vec{p}\cdot \vec{x})}|_{p^0=|\vec{p}|}$,
with the vacuum state equal to  $|\;\,1>.$
}.

%
\begin{eqnarray}
{\cal A}\,  &=& \int \; d^4x \;\frac{1}{2}\, 
(\bar{\psi} \, \gamma^{\alpha} p_{0\alpha} \psi)
+ h.c. +
\nonumber\\
& & \int \; d^4x \;\sum_{i=(I,II)}\,
{}^{i}{\hat F}^{m\,f}_{\alpha \beta}\;\, {}^{i}{\hat F}^{m f \alpha \beta}\,,
%
%
\nonumber\\
p_{0\alpha}  &=& p_{\alpha}  -
\sum_{m f}   {}^{I}{ \hat {\cal A}}^{m \dagger}_{f}\,\,
{}^{I}{\cal C}^{m}_{f \alpha}   -
\sum_{m f} {}^{II}{\hat {\cal A}}^{m \dagger}_{f}\,\,
{}^{II}{\cal C}^{m}_{f \alpha}\,,
\nonumber\\
{}^{i}{\hat F}^{m\,f}_{\alpha \beta}&=&  \partial_{\alpha}
{}^{i}{\hat{\cal A}}^{m \dagger}_{f \beta} - \partial_{\beta}
{}^{i}{\hat{\cal A}}^{m \dagger}_{f \alpha} + \varepsilon
{\bf f}^{m f m' f\,` m'' f\,''  } \,\,
{}^{i}{\hat{\cal A}}^{m' \dagger}_{f\,' \alpha}
\,{}^{i}{\hat{\cal A}}^{m'' \dagger}_{f\,'' \beta} \,,\nonumber\\
&& i=(I,II) \,.
%
\label{wholeaction}
\end{eqnarray}
$\psi$ represent $2^{\frac{d}{2}-1}$ members of one family of all 
the $2^{\frac{d}{2}-1}$ families, and $\bar{\psi} =(\psi)^{\dagger} \gamma^{0}$ their $2^{\frac{d}{2}-1}\times 2^{\frac{d}{2}-1} $ 
Hermitian conjugated partners (multiplied by $\gamma^{0}$). 
${}^{i}{\hat {\cal A}}^{m \dagger}_{f}\,\,{}^{i}{\cal C}^{m}_{f \alpha}, i=(I,II),$ represent two orthogonal groups of ``basis vectors'' in a 
tensor product with basis in ordinary space-time, extended by the space 
index  $\alpha=\mu$ for vectors  and $\alpha \ge 5$ for scalars in ordinary 
space-time, each with $2^{\frac{d}{2}-1}\times$ $2^{\frac{d}{2}-1} $ members~\footnote{The assumption that
all the fermion and boson fields are active (have non zero 
momentum) only in $d=(3+1)$ ordinary space-time mean that the 
derivative with respect to $x_{\alpha}$ gives zero.
}. The assumed action, Eq.~(\ref{wholeaction}), needs further studies,
since the studies so far were mainly done by  Eqs.~(100, 101) of the 
reference~\cite{nh2021RPPNP}.

Let us repeat:  The description of the internal spaces of fermion
and boson fields with the odd and even Clifford algebra objects,
respectively, offers an equal number of fermions and bosons,
demonstrating a (kind of) supersymmetry. However, none of
the ``basis vectors'' of the boson fields, obtained by the
application of $\gamma^a$ on the fermion ``basis vectors''
(the application from the left-hand side generates
${}^{II}{\hat {\cal A}}^{m \dagger}_{f}$, the application
from the right-hand side generates
${}^{I}{\hat {\cal A}}^{m \dagger}_{f}$) do carry the
same internal charges as  the fermion ``basis
vectors''  ~\footnote{%
The quantum numbers of $ u_{L}^{c1}$, presented in
Table~\ref{Table so13+1.} in the seventh line
$ u_{L}^{c1}$$(\equiv \stackrel{03}{[-i]}\,\stackrel{12}{[+]}|
\stackrel{56}{[+]}\,\stackrel{78}{[-]}||\stackrel{9 \;10}{(+)}
\;\;\stackrel{11\;12}{[-]}\;\;\stackrel{13\;14}{[-]}$, are:
$S^{12}$ $=\frac{1}{2}$, $S^{03}$ $=-\frac{i}{2}$,
$\tau^{13}=\frac{1}{2}(S^{56}-S^{78})=
\frac{1}{2}$, $\tau^{23}=\frac{1}{2}(S^{56}+S^{78})=0$,
$\tau^{33}=\frac{1}{2}(S^{9\, 10}- S^{11\,12})=
\frac{1}{2}$, $\tau^{38}=\frac{1}{2\sqrt{3}}(S^{9\, 10}+
S^{11\,12}- 2 S^{13\,14})=\frac{1}{2\sqrt{3}}$.\\
For the corresponding boson ``basic vector'' obtained by
multiplication of $ u_{L}^{c1}$ from the left-hand side
by $\gamma^9$ one obtains: $\gamma^9$ $ u_{L}^{c1}
(\equiv
\stackrel{03}{[-i]}\,\stackrel{12}{[+]}|\stackrel{56}{[+]}
\,\stackrel{78}{[-]}||\stackrel{9 \;10}{(+)}
\;\;\stackrel{11\;12}{[-]}\;\;\stackrel{13\;14}{[-]}$) leading
to $\stackrel{03}{[-i]}\,\stackrel{12}{[+]}|\stackrel{56}{[+]}
\,\stackrel{78}{[-]}||\stackrel{9 \;10}{[-]}
\;\;\stackrel{11\;12}{[-]}\;\;\stackrel{13\;14}{[-]}$.
Having all the quantum numbers equal zero, this boson
represents a photon. Having all the members of the algebraic
product equal to projectors, and taking into account that for
even ``basis vectors'' ${\cal S}^{ab}= (S^{ab} +
\tilde{S}^{ab})$, which gives zero for the projectors, the
product of projectors only means that such an object can not
change the internal space of odd ``basis vectors'',
what photons do not. Photons can give to fermions only the
momentum in the ordinary space time, Eq.~(\ref{phAqeqe}).
}.

We represent in Sect.~\ref{basisvectors0} the ``basis
vectors'' of fermion and boson fields for any $d=2(2n+1)$,
in particular also for $d=(13+1)$~\footnote{%
Let us repeat that the choice $d=(13+1)$
offers to describe all the quarks and antiquarks and
leptons and antileptons observed so far, predicts the existence
of the right-handed neutrinos and left-handed antineutrinos
and another weak field; requires the existence of families
of quarks and leptons and the existence of the dark matter;
announces the existence of a fourth family to the observed
three; explains the existence of the scalar fields as the
Higgs boson; predicts new scalar fields that gave rise to
inflation of the universe after the Big Bang, offers
the explanation for many a cosmological observation.}.

We demonstrated the ``basis vectors'' for the
corresponding vector and scalar gauge fields observed so
far, Subsect.~\ref{d=(13+1)}: Photons, gravitons, weak
bosons and gluons.~\footnote{%
There are additional gauge fields which can
not be observed since this would break Lorentz invariance
if requiring Lorentz transformations of the kind $M^{as},
a=(0,1,2,3)$ and $s\ge 5$ due to  the assumption that
all the fields have non zero momentum only in $d=(3+1)$.
Without  additional break of symmetry there could exist
gauge fields, which change more than only one kind of charge
at the same time.}

We can obtain gravitino if we extend points in ordinary
space-time to strings.
Describing strings with ``basis vectors'' being the eigenvectors
of the Cartan subalgebra members $S^{01}, {\tilde S}^{01},
{\cal S}^{01}=(S^{01} + {\tilde S}^{01})$, on a string
($\sigma,\tau$), we have, in this case, two odd --- 
one ``basis vector'' and one Hermitian conjugated partner ---
and two even objects,
Eq.~(\ref{1+1oddeven}),
\begin{small}
\begin{eqnarray}
\label{1+1oddeven1}
&& {\rm  \; \;odd}\nonumber\\
\hat{b}^{ 1 \dagger}_{1s}&=&\stackrel{01}{(+i)_s}\,, \quad
\hat{b}^{ 1 }_{1s}=\stackrel{01}{(-i)}\,,\nonumber\\
&&{\rm \; \;even} \;\nonumber\\
{}^{I}{\bf {\cal A}}^{1 \dagger}_{1 s}&=&\stackrel{01}{[+i]_s}\,,
\quad
{}^{II}{\bf {\cal A}}^{1 \dagger}_{1s}=\stackrel{01}{[-i]_s}\,.
\nonumber
\end{eqnarray}
\end{small}
Index $s$ points out that these ``basis vectors'' belong to the 
strings extensions and not to the ``basis vectors'' representing
internal spaces of fermion and boson fields.
The two odd ``basis vectors'' in the above equation are
Hermitian conjugated to each other. Let us make a choice that
$\hat{b}^{ 1 \dagger}_{1 s} \equiv \stackrel{01}{(+i)_s}$ is 
the ``basis vector'' (applying on the vacuum state, 
Eq.~(\ref{vaccliffodd}), $|\psi_{oc s}>=
\stackrel{01}{[-i]_{s}}$). There is only one family
($2^{\frac{d}{2}-1}=1$) with one member.
The eigenvalue $S^{01}$ of $\hat{b}^{ 1 \dagger}_{1 s} 
(= \stackrel{01}{(+i)_s})$ is $\frac{i}{2}$.

Each of the two even ``basis vectors'' is self adjoint
($({}^{I,II}{\bf {\cal A}}^{1 \dagger}_{1 s})^{\dagger}=$
${}^{I,II}{\bf {\cal A}}^{1 \dagger}_{1 s}$), with the
eigenvalues ${\cal S}^{01}$ equal  to $0$.

The internal space of photons \\(like it is
${}^{I}{\hat{\cal A}}^{\dagger}_{ph\,\bar{u}^{\bar{c1}}_{R}
\rightarrow \bar{u}^{\bar{c1}}_{R}}
(\equiv \stackrel{03}{[+i]}\stackrel{12}{[+]} \stackrel{56}{[-]}
\stackrel{7 8}{[+]}\stackrel{9\,10}{[-]} \stackrel{11\,12}{[+]}
\stackrel{13\,14}{[+]})$, Eq.(\ref{phAqeqe})), extended by a 
tensor product,
$\, *_{T \, '}\,$, with a string $\hat{b}^{1\dagger}_{1 s} (\equiv
\stackrel{01}{(+i)_s}$), can represent anticommuting
gravitinos since the photon carries the space-time
index $\mu$. 

The extensions of all the other ``basis vectors''  of space-time --- 
either the odd ones describing the internal spaces of 
fermions, or even ones describing the internal spaces of 
bosons ---  by the tensor product, $\, *_{T \, '}\,$, with the two self adjoint 
``basis vectors'' describing the internal space on the string,
$ {}^{i}{\bf {\cal A}}^{1 \dagger}_{1 s}, i=(I,II)$, do not change
commutation properties. The extended ``basis vectors'' remain 
commuting or anti-commuting. 

The extensions of  ``basis vectors''  of space-time by the tensor 
product, $\, *_{T \, '}\,$, with $\hat{b}^{ 1 \dagger}_{1 s}$ do change 
the commutation relations: The commuting ones become 
anti-commuting, the anti-commuting become commuting.

The meaning of the $2^{\frac{14}{2}-1}$ ``basis vectors'' in
tensor extension by $\hat{b}^{ 1 \dagger}_{1 s}$ needs
further study.

The extension of point fields in $d=(3+1)$ ordinary space-time (with
``basis vectors'' determined in $d=(13+1)$) to strings needs the
corresponding change of the action, presented in
Eq.~(\ref{wholeaction}), which is under consideration.

In a separate paper of this proceedings  the author discusses 
the content of the ``basis vectors'' after the ``basis vectors'' 
describing the internal spaces of fermions
($\hat{b}^{m \dagger}_{f}$) and 
bosons (${}^{I,II}{\bf {\cal A}}^{m \dagger}_{f}$) in 
$d=(13+1)$ are extended by a tensor products with the string's 
``basis vectors'' of fermions  ($\hat{b}^{1 \dagger}_{1s}$) 
and bosons (${}^{I,II}{\bf {\cal A}}^{1 \dagger}_{1s}$) 
presented in Eq.~(\ref{1+1oddeven}). 

It follows:\\
$$\hat{b}^{m \dagger}_{f}\,*_{T\, '}\,
{}^{I,II}{\bf {\cal A}}^{1 \dagger}_{1s}$$ 
represent the anti-commuting ``basis vectors'' of fermions in
the internal space $SO(d-1, 1)$, having the Hermitian 
conjugated partners 
in a separate group; the projectors $\stackrel{01}{[\pm i]_s}$
of ${}^{I,II}{\bf {\cal A}}^{1 \dagger}_{1s}$, 
namely, do not change anti-commutativity of fermion ``basis 
vectors''.\\
$${}^{I,II}{\bf {\cal A}}^{m \dagger}_{f}\,*_{T\,'}\,
{}^{I,II}{\bf {\cal A}}^{1 \dagger}_{1s}$$
represent the commuting ``basis vectors'' of bosons in the
internal space $SO(d-1, 1)$, appearing in two orthogonal 
groups; the projectors $\stackrel{01}{[\pm i]_s}$ of 
${}^{I,II}{\bf {\cal A}}^{1 \dagger}_{1s}$, namely, do 
not change commutativity of boson ``basis vectors''.\\ 
$${}^{I,II}{\bf {\cal A}}^{m \dagger}_{f}\,*_{T\,'}\,
\hat{b}^{1 \dagger}_{1s}$$
represent the anti-commuting ``basis vectors'' of new 
fermions, with the quantum numbers of bosons in the internal 
space $SO(d-1, 1)$;
the nilpotent $\hat{b}^{ 1 \dagger}_{1 s}
(\equiv \stackrel{01}{(+i)_s})$ on the string, namely, 
changes commutativity 
of boson ``basis vectors'' keeping their quantum numbers 
unchanged.\\ 
$$\hat{b}^{m \dagger}_{f}\,*_{T\, '}\,
\hat{b}^{1 \dagger}_{1s}$$ 
represent the commuting ``basis vectors'' of new bosons  
with the quantum numbers of fermions in the internal space 
$SO(d-1, 1)$, $\hat{b}^{ 1 \dagger}_{1 s}$ $(\equiv 
\stackrel{01}{(+i)}) $ change the anti-commutativity of 
$\hat{b}^{m \dagger}_{f}$.\\

This does not look as the usually desired supersymmetry,
requiring that each fermion (in our case one of 
$2^{\frac{d}{2}-1}$ members appearing in 
$2^{\frac{d}{2}-1}$ families, having their Hermitian 
conjugated partner in a separate group) has a 
supersymmetric  partner with the same charges and  with
the spin $0$. It does not look either that any boson field (in 
our case any one of two groups each with 
$2^{\frac{d}{2}-1}\times$$2^{\frac{d}{2}-1}$ members)
has a supersymmetric partner with  the same charges and 
with the spin $\frac{1}{2}$.\\


There is another possibility to achieve the supersymmetric 
partners to the ``basis vectors'' presented fermions and 
bosons in $2(2n+1)$ dimensional internal spaces of
fermions and bosons. It will be discussed in a separate 
contribution to the Bled Proceeding 2024. One can namely 
go to an odd dimensional space $d=2(2n+1) +1$.
As discussed in the article~\cite{n2023MDPI} there are 
two groups of ``basis vectors'' in odd dimensional spaces.\\


One group determines the anti-commuting ``basis vectors'' of 
$2^{\frac{d-1}{2}-1}$ fermions appearing in 
$2^{\frac{d-1}{2}-1}$ families with their $2^{\frac{d}{2}-1}$
$\times2^{\frac{d}{2}-1}$ Hermitian conjugated partners 
appearing in a separate group, as well as two 
orthogonal groups each with $2^{\frac{d}{2}-1}$
$\times2^{\frac{d}{2}-1}$ of ``basis vectors'',  with their
Hermitian conjugated partners within the same group.

The second group determines anti-commuting ``basis 
vectors''  appearing in two separate orthogonal groups each 
with $2^{\frac{d}{2}-1}$ $\times2^{\frac{d}{2}-1}$ of 
``basis vectors'',  with their Hermitian conjugated 
partners within the same group, as well as the commuting 
``basis vectors'' of ``fermions'' appearing in families with 
their Hermitian conjugated partners in a separate group.\\

Also these two groups might demonstrate a kind of 
supersymmetry.
 
 


 
 


Let  us finish this section by pointing out  that the proposed theory 
offers:\\
Photons with ``basis vectors'' including only projectors;
gravitons with ``basis vectors'' having two nilpotents only in 
$SO(3,1)$ of $SO(13,1)$, all the rest are projectors; weak 
bosons with  ''basis vectors'' which have two nilpotents only 
in $SO(4)$ manifesting as $SU(2)\times SU(2)$ of $SO(13,1)$; 
and gluons ``basis vectors'' with two nilpotents in 
$SU(3) \times U(1)$ of $SO(13,1)$, all the rest are projectors.

All the observed gauge fields, either vectors or scalars, are 
presented by only two nilpotents all the rest are projectors.
There are  breaks of symmetries~\cite{nh2021RPPNP} which
make choices.\\

The extension of the ``basis vectors'' to strings in the context 
of renormalizability, as well as the offer of odd-dimensional spaces 
needs further studies.
\section{Conclusions}

\label{conclusions}
This contribution presents the ``basis vectors'' of internal spaces of
fermion and boson second quantized fields, Sect.~\ref{basisvectors0},
manifesting that fermion ``basis vectors'', written as products of an
odd number of nilpotents (at least one, the rest of projectors),
anti-commute, while boson ``basis vectors'', written as products of
an even number of nilpotents (or of only projectors), commute,
explaining the second quantization postulates of Dirac for fermions
and bosons.

Each nilpotent and projector is chosen to be the eigenvector of one
of the members of the Cartan subalgebra, Eq.~\ref{cartangrasscliff};
correspondingly are the ``basis vectors'' eigenvectors of all the
Cartan subalgebra members; they offer a transparent and elegant
way to find  the Dirac matrices in any even dimensional space and
the corresponding matrices for the boson fields~\footnote{%
The usual presentation of Dirac matrices in higher dimensional
spaces is much more complicated and much less transparent.
The same is true for the usual group presentations
for either fermions or bosons. The Dirac matrices in $d=(3+1)$
are designed for massive fermions, they do not anticommute.
Dirac, having no $\tilde{\gamma}^{a}$, does not include families
of fermions; although he could, since  he has $2^d$ products of
odd and even possibilities of products of $\gamma^{a}$.}. 
However, knowing the states and operators, ${\cal S}^{ab},
\tilde{S}^{ab}, {\cal S}^{ab}=(\tilde{S}^{ab}+S^{ab}),
{}^{I}{\hat{\cal A}}^{m \dagger}_{f}, 
{}^{II}{\hat{\cal A}}^{m \dagger}_{f}$, the matrix 
representation is not even needed.

We demonstrated that the number of fermion ``basis vectors''
and their Hermitian conjugated partners are equal to the number
of ``basis vectors'' of bosons, appearing in two orthogonal
groups, manifesting a (kind of) supersymmetry.

Representing the corresponding creation and annihilation 
operators for fermion and boson fields as tensor products, 
$\,*_T\,$, of ``basis vectors'' and basis in ordinary 
space-time, Eqs.~(\ref{wholespacefermions}, 
\ref{wholespacebosons}), the boson fields have to obtain the 
space index so that the creation and annihilation operators manifest 
properties of the second quantized anti-commuting fermion 
fields and commuting boson fields, 
Sect.~\ref{creationannihilation}.

We assumed in this talk that fermions and bosons have non 
zero momenta only in $d=(3+1)$, while the involved internal 
space needs to be $d=(13+1)$ to be able to offer the observed
properties of quarks and leptons and antiquarks and antileptons,
appearing in families, as well as the corresponding vector (if the
space index $\alpha$ is ($0,1,2,3$)) and scalar (if the space index 
$\alpha$ is $\ge 5$) boson fields.

We presented ``basis vectors'' for fermion and boson fields in
any $d=2(2n+1)$,  Sect.~\ref{basisvectors0}, in particular in
$d=(1+1)$ while searching for the ``basis vectors'' for the
strings, for $d=(5+1)$ while manifesting how do the ``basis
vectors'' work in a simple case with fermions representing only
``positrons'' and ``electrons'' appearing in families and with
bosons representing ``photons'' and ``gravitons'', and in
$d=(13+1)$ manifesting ``basis vectors'' which explain the
appearance of the observed quarks and leptons and antiquarks
and antileptons appearing in families and of  photons, weak
bosons, gluons, and a second kind of weak bosons (not yet observed)
and gravitons (not yet observed),
Subsect.~\ref{bosons13+1and5+1}. There are two kinds of
boson ``basis vectors''; one kind transforms the family
members of any family among themselves, the second kind
transforms any of the family members into the same family
member of all the families.


Does the description of the second quantized fermion and boson
fields with the ``basis vectors'', bringing a new understanding
of the second quantized fermion and boson fields, bring also a 
new understanding of cosmology? It is at least a very promising 
suggestion. Let us assume that at Big Bang, all the fermion, 
vector and scalar boson fields were massless, with the ``basis 
vectors''  in internal space determined by $SO(13,1)$; fermions, 
vector and scalar gauge fields have non zero momenta only in 
$d=(3+1)$ of ordinary space-time; all the scalar gauge fields 
with the space index $\alpha \ge 5$ (with non-zero momentum 
only in $d=(3+1)$) contribute to inflation.
The Lorentz transformations of the kind $M^{ms}=L^{ms} +
S^{ms}$ (or $M^{ms}=L^{ms} + {\cal S}^{ms}$),
$m=(0,1,2,3)$, $s\ge5$ are not possible (since $L^{ms}$
can not be performed).

As discussed in~(\cite{nh2021RPPNP} and the references 
therein), the condensate breaks symmetry, so that the second 
weak $SU(2)_{II}$ bosons, interacting with the
condensate, become at low energies massive; all the rest
gauge fields ($SU(2)_{I}$, $SU(3), U(1)$, 
gravity)~\cite{nh2021RPPNP}, do not interact with the
condensate, remaining therefore massless up to the electroweak 
break, which is caused by scalar fields with the space index 
$\alpha=(5,6,7,8)$. After the electroweak break there are 
photons, gluons and gravitons which keep their 
masslessness~\cite{n2024NPB}.

We demonstrate that the ``basis vectors''  for boson second
quantized fields can also be expressed as algebraic products of
fermion ``basis vectors''  and their Hermitian conjugated
partners; consequently,  we need only to know the fermion 
``basis vectors''.

In this contribution, 
the trial to extend the points in ordinary space-time to strings
is presented in Sect.~\ref{string}. The internal space of strings, 
that is their ``basis vectors'', were added in a tensor product,
$\,*_{T \, '}\,$, to the ``basis vectors'' of fermions and bosons 
describing the $d=(13+1)$ internal space. The tensor 
products of the fermion and  boson ``basis vectors'' with the 
bosons self adjoint ``basis vectors'' on the string leads to
the anti-commuting creation and annihilation operators for 
fermions and commutation operators for bosons, suggesting 
that at low energies only the internal spaces of the second 
quantized fields with the ``basis vectors''  describing the
$d=(13+1)$ internal space are important.  The  tensor 
product of the ``basis vectors'' of fermions and bosons 
describing the $d=(13+1)$ internal space with the odd
basis vectors on a string gives to the commuting ``basis 
vectors'' in $d=(13+1)$ the anti-commuting partners with the
same charges, and to the anti-commuting ``basis 
vectors'' in $d=(13+1)$ the commuting partners with the
same charges. This, it might be an interesting part, needs 
further studies.

There is another possibility to achieve the supersymmetric 
partners to the ``basis vectors'' presented fermions and 
bosons in $2(2n+1)$-dimensional internal spaces of
fermions and bosons, Sect.~\ref{string}. 
 
 In an odd dimensional space $d=2(2n+1) +1$  there are 
two groups of ``basis vectors''~\cite{n2023MDPI}:
One group determines the anti-commuting ``basis vectors'' of 
$2^{\frac{d-1}{2}-1}$ fermions appearing in 
$2^{\frac{d-1}{2}-1}$ families, with their $2^{\frac{d-1}{2}-1}$
$\times2^{\frac{d-1}{2}-1}$ Hermitian conjugated partners 
appearing in a separate group, as well as two 
orthogonal groups each with $2^{\frac{d-1}{2}-1}$
$\times2^{\frac{d-1}{2}-1}$ of ``basis vectors'',  with their
Hermitian conjugated partners within the same group.

The second group determines anti-commuting ``basis 
vectors''  appearing in two separate orthogonal groups each 
with $2^{\frac{d-1}{2}-1}$ $\times2^{\frac{d-1}{2}-1}$ of 
``basis vectors'',  with their Hermitian conjugated 
partners within the same group, as well as the commuting 
``basis vectors'' of ``fermions'' appearing in families with 
their Hermitian conjugated partners in a separate group.

Also these two groups might demonstrate a kind of 
supersymmetry, suggesting to be used to achieve 
renormalizability.

We want to understand whether the elegant and simple
description of the internal degrees of freedom of fermions and
bosons with the odd and even ``basis vectors''
manifesting a kind of supersymmetry --- offering explanations
for so many observed properties of fermion and boson second
quantized fields, explaining as well the second quantization
postulates for fermion and boson fields, offering expressions
for boson ``basis vectors'' (their internal spaces) as algebraic
products of fermion ``basis vectors'' and their Hermitian
conjugated partners, offering explanations for cosmological
observations with inflation included --- can be related to strings,
or it offers a new way to understand renormalizability on all levels
of energy.

\appendix
\section{Useful tables}
\label{tables}

These Tables are mainly taken from Refs.~\cite{n2023NPB,n2024NPB} and
are meant to illustrate Sects.~(\ref{basisvectors0},\ref{FermionsVetorScalar}
\ref{bosons13+1and5+1},\ref{relationsCliffordoddeven}).

The case with $d=(5+1)$ is meant to learn how odd and 
even ``basis vectors'' illustrate internal spaces of fermions and bosons,
Table~\ref{Table Clifffourplet.}.

Fermions, in this case ``electrons'' with a negative ``charge'', $S^{56}=
- \frac{1}{2}$, and ``positrons'' with a positive ``charge'', $S^{56}=
\frac{1}{2}$, appear in $16$ odd  ``basis vectors'' (with one or three
nilpotents), in $4$ families with $4$ members in each family, having
$16$ Hermitian conjugated partners.

Bosons, in this case ``photons'' and  ``gravitons'', appear in 
two orthogonal groups; each group has $16$ members, and their 
Hermitian conjugated partners appear within the same group. 
The eigenvalues of the
Cartan subalgebra members ${\cal S}^{ab}=(S^{ab}+ \tilde{S}^{ab})$
are equal to either $(\pm i,0)$ or $(\pm 1,0)$. ``Photons'' are products of
only projectors, ``gravitons'' have two nilpotents with ${\cal S}^{03}$
equal to $(\pm i)$ and ${\cal S}^{12}$ equal to $(\pm 1)$.

Tables~(\ref{S120Cliff basis5+1even I.},
\ref{transverseCliff basis5+1even I.}), manifest that the even
``basis vectors'' can be expressed as products of the odd ``basis
vectors'' and their Hermitian conjugated partners. Let us remind 
the reader that projectors are self adjoint operators, while the Hermitian
conjugated partner to $\stackrel{ab}{(\pm i)}$ are $\stackrel{ab}{(\mp i)}$
and  to $\stackrel{ab}{(\pm 1)}$ are $\stackrel{ab}{(\mp 1)}$,
Eq.~(\ref{usefulrel0}).

\begin{table*}
\begin{small}
\caption{\label{Table Clifffourplet.}  This table, taken 
from~\cite{n2023NPB}, represents $2^d=64$ ``basis vectors", which 
are the eigenstates of the Cartan subalgebra members, 
Eq.~(\ref{cartangrasscliff}). Half of them are the odd ``basis 
vectors" with the odd number of nilpotents (one or three), the other 
half are the even ``basis vectors'' with the even number of 
nilpotents (two or none). They are divided into four groups. The first 
group, $odd \,I$, represents odd ``basis vectors'' 
${\hat b}^{m \dagger}_f$, appearing 
in $2^{\frac{d}{2}-1}=4$ families ($f=1,2,3,4$), each family 
having $2^{\frac{d}{2}-1}=4$ family members ($m=1,2,3,4$).
The second group, $odd\,II$, contains Hermitian conjugated partners 
of the first group (for each family separately), ${\hat b}^{m}_f=$
$({\hat b}^{m \dagger}_f)^{\dagger}$. Either $odd \,I$ or $odd \,II$ 
are products of an odd number of nilpotents (one or three) and of 
projectors (two or none).
The family quantum numbers of ${\hat b}^{m \dagger}_f$ (the 
eigenvalues of $(\tilde{S}^{03}, \tilde{S}^{12},\tilde{S}^{56})$) 
appear for the first {\it odd I } group above each family, the quantum
numbers of the ``family'' members $(S^{03}, S^{12}, S^{56})$ are
written in the last three columns. 
For the Hermitian conjugated partners of {\it odd I}, presented in the 
group {\it odd II}, the quantum numbers $(S^{03}, S^{12}, S^{56})$ 
are presented above each group of the Hermitian conjugated partners, 
the last three columns tell eigenvalues of $(\tilde{S}^{03}, 
\tilde{S}^{12},\tilde{S}^{56})$.
The two groups with the even number of nilpotents (two or none), 
{\it even \,I} and {\it even \,II}, have their Hermitian conjugated 
partners within its groups. The quantum numbers $f$, that is the 
eigenvalues of $(\tilde{S}^{03}, \tilde{S}^{12},\tilde{S}^{56})$, 
are written above column of four members, the quantum numbers 
of the members, $(S^{03}, S^{12}, S^{56})$, are written in the 
last three columns. The quantum numbers of the even ``basis 
vectors'' are $({\cal {\bf S}}^{03}, {\cal {\bf S}}^{12}, 
{\cal {\bf S}}^{56})$, determined by ${\cal {\bf S}}^{ab}$ 
$= (S^{ab} + \tilde{S}^{ab} )$.
 \vspace{2mm}}
 \end{small}
\begin{tiny}
\begin{center}
  \begin{tabular}{|c|c|c|c|c|c|r|r|r|}
\hline
$ $&$$&$ $&$ $&$ $&&$$&$$&$$\\
$''basis\, vectors'' $&$m$&$ f=1$&$ f=2 $&$ f=3 $&
$ f=4 $&$$&$$&$$\\ 
$(\tilde{S}^{03}, \tilde{S}^{12}, \tilde{S}^{56})$&$\rightarrow$&$(\frac{i}{2},- \frac{1}{2},-\frac{1}{2})$&$(-\frac{i}{2},-\frac{1}{2},\frac{1}{2})$&
$(-\frac{i}{2},\frac{1}{2},-\frac{1}{2})$&$(\frac{i}{2},\frac{1}{2},\frac{1}{2})$&$S^{03}$
 &$S^{12}$&$S^{56}$\\ 
\hline
$ $&$$&$ $&$ $&$ $&&$$&$$&$$\\ 
$odd \,I\; {\hat b}^{m \dagger}_f$&$1$& 
$\stackrel{03}{(+i)}\stackrel{12}{[+]}\stackrel{56}{[+]}$&
                        $\stackrel{03}{[+i]}\stackrel{12}{[+]}\stackrel{56}{(+)}$ & 
                        $\stackrel{03}{[+i]}\stackrel{12}{(+)}\stackrel{56}{[+]}$ &  
                        $\stackrel{03}{(+i)}\stackrel{12}{(+)}\stackrel{56}{(+)}$ &
                        $\frac{i}{2}$&$\frac{1}{2}$&$\frac{1}{2}$\\ 
$$&$2$&    $[-i](-)[+] $ & $(-i)(-)(+) $ & $(-i)[-][+] $ & $[-i][-](+) $ &$-\frac{i}{2}$&
$-\frac{1}{2}$&$\frac{1}{2}$\\ 
$$&$3$&    $[-i] [+](-)$ & $(-i)[+][-] $ & $(-i)(+)(-) $ & $[-i](+)[-] $&$-\frac{i}{2}$&
$\frac{1}{2}$&$-\frac{1}{2}$\\ 
$$&$4$&    $(+i)(-)(-)$ & $[+i](-)[-] $ & $[+i][-](-) $ & $(+i)[-][-]$&$\frac{i}{2}$&
$-\frac{1}{2}$&$-\frac{1}{2}$\\ 
\hline
$ $&$$&$ $&$ $&$ $&&$$&$$&$$\\ 
$(S^{03}, S^{12}, S^{56})$&$\rightarrow$&$(-\frac{i}{2}, \frac{1}{2},\frac{1}{2})$&
$(\frac{i}{2},\frac{1}{2},-\frac{1}{2})$&
$(\frac{i}{2},- \frac{1}{2},\frac{1}{2})$&$(-\frac{i}{2},-\frac{1}{2},-\frac{1}{2})$&
$\tilde{S}^{03}$
&$\tilde{S}^{12}$&$\tilde{S}^{56}$\\ 
&&
$\stackrel{03}{\;\,}\;\;\,\stackrel{12}{\;\,}\;\;\,\stackrel{56}{\;\,}$&
$\stackrel{03}{\;\,}\;\;\,\stackrel{12}{\;\,}\;\;\,\stackrel{56}{\;\,}$&
$\stackrel{03}{\;\,}\;\;\,\stackrel{12}{\;\,}\;\;\,\stackrel{56}{\;\,}$&
$\stackrel{03}{\;\,}\;\;\,\stackrel{12}{\;\,}\;\;\,\stackrel{56}{\;\,}$&
&&\\
\hline
$ $&$$&$ $&$ $&$ $&&$$&$$&$$\\ 
$odd\,II\; {\hat b}^{m}_f$&$1$ &$(-i)[+][+]$ & $[+i][+](-)$ & $[+i](-)[+]$ & $(-i)(-)(-)$&
$-\frac{i}{2}$&$-\frac{1}{2}$&$-\frac{1}{2}$\\ 
$$&$2$&$[-i](+)[+]$ & $(+i)(+)(-)$ & $(+i)[-][+]$ & $[-i][-](-)$&
$\frac{i}{2}$&$\frac{1}{2}$&$-\frac{1}{2}$\\ 
$$&$3$&$[-i][+](+)$ & $(+i)[+][-]$ & $(+i)(-)(+)$ & $[-i](-)[-]$&
$\frac{i}{2}$&$-\frac{1}{2}$&$\frac{1}{2}$\\ 
$$&$4$&$(-i)(+)(+)$ & $[+i](+)[-]$ & $[+i][-](+)$ & $(-i)[-][-]$&
$-\frac{i}{2}$&$\frac{1}{2}$&$\frac{1}{2}$\\ 
\hline
&&&&&&&&\\ 
\hline
$ $&$$&$ $&$ $&$ $&&$$&$$&$$\\ 
$(\tilde{S}^{03}, \tilde{S}^{12}, \tilde{S}^{56})$&$\rightarrow$&
$(-\frac{i}{2},\frac{1}{2},\frac{1}{2})$&$(\frac{i}{2},-\frac{1}{2},\frac{1}{2})$&
$(-\frac{i}{2},-\frac{1}{2},-\frac{1}{2})$&$(\frac{i}{2},\frac{1}{2},-\frac{1}{2})$&
$S^{03}$&$S^{12}$&$S^{56}$\\ 
&& 
$\stackrel{03}{\;\,}\;\;\,\stackrel{12}{\;\,}\;\;\,\stackrel{56}{\;\,}$&
$\stackrel{03}{\;\,}\;\;\,\stackrel{12}{\;\,}\;\;\,\stackrel{56}{\;\,}$&

$\stackrel{03}{\;\,}\;\;\,\stackrel{12}{\;\,}\;\;\,\stackrel{56}{\;\,}$&
$\stackrel{03}{\;\,}\;\;\,\stackrel{12}{\;\,}\;\;\,\stackrel{56}{\;\,}$&
&&\\ 
\hline
$ $&$$&$ $&$ $&$ $&&$$&$$&$$\\ 
$even\,I \; {}^{I}{\cal A}^{m}_f$&$1$&$[+i](+)(+) $ & $(+i)[+](+) $ & $[+i][+][+] $ & $(+i)(+)[+] $ &$\frac{i}{2}$&
$\frac{1}{2}$&$\frac{1}{2}$\\ 
$$&$2$&$(-i)[-](+) $ & $[-i](-)(+) $ & $(-i)(-)[+] $ & $[-i][-][+] $ &$-\frac{i}{2}$&
$-\frac{1}{2}$&$\frac{1}{2}$\\ 
$$&$3$&$(-i)(+)[-] $ & $[-i][+][-] $ & $(-i)[+](-) $ & $[-i](+)(-) $&$-\frac{i}{2}$&
$\frac{1}{2}$&$-\frac{1}{2}$\\ 
$$&$4$&$[+i][-][-] $ & $(+i)(-)[-] $ & $[+i](-)(-) $ & $(+i)[-](-) $&$\frac{i}{2}$&
$-\frac{1}{2}$&$-\frac{1}{2}$\\ 
\hline
$ $&$$&$ $&$ $&$ $&&$$&$$&$$\\ 
$(\tilde{S}^{03}, \tilde{S}^{12}, \tilde{S}^{56})$&$\rightarrow$&
$(\frac{i}{2},\frac{1}{2},\frac{1}{2})$&$(-\frac{i}{2},-\frac{1}{2},\frac{1}{2})$&
$(\frac{i}{2},-\frac{1}{2},-\frac{1}{2})$&$(-\frac{i}{2},\frac{1}{2},-\frac{1}{2})$&
$S^{03}$&$S^{12}$&$S^{56}$\\ 
&& 
$\stackrel{03}{\;\,}\;\;\,\stackrel{12}{\;\,}\;\;\,\stackrel{56}{\;\,}$&
$\stackrel{03}{\;\,}\;\;\,\stackrel{12}{\;\,}\;\;\,\stackrel{56}{\;\,}$&
$\stackrel{03}{\;\,}\;\;\,\stackrel{12}{\;\,}\;\;\,\stackrel{56}{\;\,}$&
$\stackrel{03}{\;\,}\;\;\,\stackrel{12}{\;\,}\;\;\,\stackrel{56}{\;\,}$&
&&\\ 
\hline
$ $&$$&$ $&$ $&$ $&&$$&$$&$$\\ 
$even\,II \; {}^{II}{\cal A}^{m}_f$&$1$& $[-i](+)(+) $ & $(-i)[+](+) $ & $[-i][+][+] $ & 
$(-i)(+)[+] $ &$-\frac{i}{2}$&
$\frac{1}{2}$&$\frac{1}{2}$\\ 
$$&$2$&    $(+i)[-](+) $ & $[+i](-)(+) $ & $(+i)(-)[+] $ & $[+i][-][+] $ &$\frac{i}{2}$&
$-\frac{1}{2}$&$\frac{1}{2}$ \\ 
$$&$3$&    $(+i)(+)[-] $ & $[+i][+][-] $ & $(+i)[+](-) $ & $[+i](+)(-) $&$\frac{i}{2}$&
$\frac{1}{2}$&$-\frac{1}{2}$\\ 
$$&$4$&    $[-i][-][-] $ & $(-i)(-)[-] $ & $[-i](-)(-) $ & $(-i)[-](-) $&$-\frac{i}{2}$&
$-\frac{1}{2}$&$-\frac{1}{2}$\\ 
\hline
 \end{tabular}
\end{center}
\end{tiny}
\end{table*}
%
%
%
\begin{table}
\begin{tiny}
\caption{ This table is taken from Ref.~\cite{n2024NPB}.
The even ``basis vectors'', ${}^{I}{\hat{\cal A}}^{m \dagger}_{f}$,
belonging to zero momentum in internal space, ${\cal S}^{12}=$ $0$,
for $d=(5+1)$, are presented as algebraic products of the $f=1$ family
``basis vectors'' $\hat{b}^{m' \dagger}_{1}$ and their Hermitian conjugated
partners ($\hat{b}^{m'' \dagger}_{1})^{\dagger}$, presented in 
Table~\ref{Table Clifffourplet.}:
$\hat{b}^{m' \dagger}_{1} *_{A}$ ($\hat{b}^{m'' \dagger}_{1})^{\dagger}$.
The two ${}^{I}{\hat{\cal A}}^{m \dagger}_{f}$ which are Hermitian
conjugated partners, are marked with the same symbol (either $\bigtriangleup$ or $\bullet$). The symbol $\bigcirc$ presents selfadjoint members.
The even ``basis vectors'' ${}^{I}{\hat{\cal A}}^{m \dagger}_{f}$
are products of one projector and two nilpotents or three projectors (they are
self-adjoint), the odd ``basis
vectors'' and their Hermitian conjugated partners are products of one nilpotent
and two projectors or of three nilpotents. The even and odd
objects are eigenvectors of all the corresponding Cartan subalgebra members,
Eq.~(\ref{cartangrasscliff}). There are $\frac{1}{2} \times 2^{\frac{6}{2}-1}
\times 2^{\frac{6}{2}-1}$ algebraic products of $\hat{b}^{m' \dagger}_{1}
*_{A}$ ($\hat{b}^{m'' \dagger}_{1})^{\dagger}$. The rest $8$ of $16$
members have ${}^{I}{\hat{\cal A}}^{m \dagger}_{f}$ with
${\cal S}^{12}=+ 1$ (four) and with ${\cal S}^{12}=- 1$ (four), presented
in Table~\ref{transverseCliff basis5+1even I.}.
The members $\hat{b}^{m' \dagger}_{f}$ together with their Hermitian
conjugated partners of each of the four families, $f=(1,2,3,4)$, offer the
same ${}^{I}{\hat{\cal A}}^{m \dagger}_{f}$ with ${\cal S}^{12}=0$
as the ones presented in this table. The table is taken from
Ref.~\cite{nh2023dec,n2024NPB}.
\vspace{3mm}}
\label{S120Cliff basis5+1even I.} 
 %
 \begin{center}
 \begin{tabular}{ |c| c| c c|}
 \hline
 $$&$$&$$&$$\\
${\cal S}^{12} $&$symbol$&${}^{I}\hat{\cal A}^{m \dagger}_f=$
&$\hat{b}^{m' \dagger}_{f `} *_A (\hat{b}^{m'' \dagger}_{f `})^{\dagger}$\\
\\
\hline
%
$$&$$&$$&$$\\
$0$&$\bigtriangleup$&${}^{I}\hat{\cal A}^{2 \dagger}_1=$&
$\hat{b}^{2 \dagger}_{1} *_{A} (\hat{b}^{4 \dagger}_{1})^{\dagger}$\\
$$&$$&$$&$$\\
$ $&$$ &$\stackrel{03}{(-i)}\,\stackrel{12}{[-]} \stackrel{56}{(+)}$&
$\stackrel{03}{[-i]}\,\stackrel{12}{(-)} \stackrel{56}{[+]} *_{A} 
\stackrel{03}{(-i)}\,\stackrel{12}{(+)} \stackrel{56}{(+)}$\\
\hline
$$&$$&$$&$$\\
$0$&$\bigcirc$&${}^{I}\hat{\cal A}^{4 \dagger}_1=$&
$\hat{b}^{4 \dagger}_{1} *_{A} (\hat{b}^{4 \dagger}_{1})^{\dagger}$\\
$$&$$&$$&$$\\
$ $&$$ &$\stackrel{03}{[+i]}\,\stackrel{12}{[-]} \stackrel{56}{[-]}$&
$\stackrel{03}{(+i)}\,\stackrel{12}{(-)} \stackrel{56}{(-)} *_{A} 
\stackrel{03}{(-i)}\,\stackrel{12}{(+)} \stackrel{56}{(+)}$\\
\hline
$$&$$&$$&$$\\
$0$&$\bullet$&${}^{I}\hat{\cal A}^{1 \dagger}_2=$&
$\hat{b}^{1 \dagger}_{1} *_{A} (\hat{b}^{3 \dagger}_{1})^{\dagger}$\\
$$&$$&$$&$$\\
$ $&$$ &$\stackrel{03}{(+i)}\,\stackrel{12}{[+]} \stackrel{56}{(+)}$&
$\stackrel{03}{(+i)}\,\stackrel{12}{[+]} \stackrel{56}{[+]} *_{A} 
\stackrel{03}{[-i]}\,\stackrel{12}{[+]} \stackrel{56}{(+)}$\\
\hline
$$&$$&$$&$$\\
$0$&$\bigcirc$&${}^{I}\hat{\cal A}^{3 \dagger}_2=$&
$\hat{b}^{3 \dagger}_{1} *_{A} (\hat{b}^{3 \dagger}_{1})^{\dagger}$\\
$$&$$&$$&$$\\
$ $&$$ &$\stackrel{03}{[-i]}\,\stackrel{12}{[+]} \stackrel{56}{[-]}$&
$\stackrel{03}{[-i]}\,\stackrel{12}{[+]} \stackrel{56}{(-)} *_{A} 
\stackrel{03}{[-i]}\,\stackrel{12}{[+]} \stackrel{56}{(+)}$\\
\hline
\hline
$$&$$&$$&$$\\
$0$&$\bigcirc$&${}^{I}\hat{\cal A}^{1 \dagger}_3=$&
$\hat{b}^{1 \dagger}_{1} *_{A} (\hat{b}^{1 \dagger}_{1})^{\dagger}$\\
$$&$$&$$&$$\\
$ $&$$ &$\stackrel{03}{[+i]}\,\stackrel{12}{[+]} \stackrel{56}{[+]}$&
$\stackrel{03}{(+i)}\,\stackrel{12}{[+]} \stackrel{56}{[+]} *_{A} 
\stackrel{03}{(-i)}\,\stackrel{12}{[+]} \stackrel{56}{[+]}$\\
\hline
$$&$$&$$&$$\\
$0$&$\bullet$&${}^{I}\hat{\cal A}^{3 \dagger}_3=$&
$\hat{b}^{3 \dagger}_{1} *_{A} (\hat{b}^{1 \dagger}_{1})^{\dagger}$\\
$$&$$&$$&$$\\
$ $&$$ &$\stackrel{03}{(-i)}\,\stackrel{12}{[+]} \stackrel{56}{(-)}$&
$\stackrel{03}{[-i]}\,\stackrel{12}{[+]} \stackrel{56}{(-)} *_{A} 
\stackrel{03}{(-i)}\,\stackrel{12}{[+]} \stackrel{56}{[+]}$\\
\hline
$$&$$&$$&$$\\
$0$&$\bigcirc$&${}^{I}\hat{\cal A}^{2 \dagger}_4=$&
$\hat{b}^{2 \dagger}_{1} *_{A} (\hat{b}^{2 \dagger}_{1})^{\dagger}$\\
$$&$$&$$&$$\\
$ $&$$ &$\stackrel{03}{[-i]}\,\stackrel{12}{[-]} \stackrel{56}{[+]}$&
$\stackrel{03}{[-i]}\,\stackrel{12}{(-)} \stackrel{56}{[+]} *_{A} 
\stackrel{03}{[-i]}\,\stackrel{12}{(+)} \stackrel{56}{[+]}$\\
\hline
$$&$$&$$&$$\\
$0$&$\bigtriangleup$&${}^{I}\hat{\cal A}^{4 \dagger}_3=$&
$\hat{b}^{4 \dagger}_{1} *_{A} (\hat{b}^{2 \dagger}_{1})^{\dagger}$\\
$$&$$&$$&$$\\
$ $&$$ &$\stackrel{03}{(+i)}\,\stackrel{12}{[-]} \stackrel{56}{(-)}$&
$\stackrel{03}{(+i)}\,\stackrel{12}{(-)} \stackrel{56}{(-)} *_{A} 
\stackrel{03}{[-i]}\,\stackrel{12}{(+)} \stackrel{56}{[+]}$\\
\hline
 \end{tabular}
 \end{center}
\end{tiny}
\end{table}
%
\begin{table}
\begin{tiny}
\caption{
The even ``basis vectors'' ${}^{I}{\hat{\cal A}}^{m \dagger}_{f}$,
belonging to transverse momentum in internal space, ${\cal S}^{12}=$
$1$, the first half of ${}^{I}{\hat{\cal A}}^{m \dagger}_{f}$, and
${\cal S}^{12}=-1$, the
second half of ${}^{I}{\hat{\cal A}}^{m \dagger}_{f}$, for $d=(5+1)$, are
presented as algebraic products of the $f=1$ family ``basis vectors''
$\hat{b}^{m' \dagger}_{1}$ and their Hermitian conjugated partners
($\hat{b}^{m'' \dagger}_{1})^{\dagger}$: $\hat{b}^{m' \dagger}_{1} *_{A}$
($\hat{b}^{m'' \dagger}_{1})^{\dagger}$. Two
${}^{I}{\hat{\cal A}}^{m \dagger}_{f}$ which are the Hermitian conjugated
partners are marked with the same symbol ($\star \star$, $\ddagger$, $\otimes$,
$\odot \odot$).
The even ``basis vectors'' ${}^{I}{\hat{\cal A}}^{m \dagger}_{f}$
are products of one projector and two nilpotents, the odd ``basis
vectors'' and their Hermitian conjugated partners are products of one nilpotent
and two projectors or of three nilpotents. The even and odd
objects are eigenvectors of all the corresponding Cartan subalgebra members,
Eq.~(\ref{cartangrasscliff}). There are $\frac{1}{2} \times 2^{\frac{6}{2}-1}
\times 2^{\frac{6}{2}-1}$ algebraic products of $\hat{b}^{m' \dagger}_{1} *_{A}$
($\hat{b}^{m'' \dagger}_{1})^{\dagger}$ with ${\cal S}^{12}$ equal
to $\pm 1$. The rest $8$ of $16$ members
present ${}^{I}{\hat{\cal A}}^{m \dagger}_{f}$ with ${\cal S}^{12}=0$.
The members $\hat{b}^{m' \dagger}_{f}$ together with their Hermitian
conjugated partners of each of the four families, $f=(1,2,3,4)$, offer the
same ${}^{I}{\hat{\cal A}}^{m \dagger}_{f}$ with ${\cal S}^{12}=\pm1$
as the ones presented in this table.
(And equivalently for ${\cal S}^{12}=0$.)
\vspace{3mm}}
\label{transverseCliff basis5+1even I.} 
 %
 \begin{center}
 \begin{tabular}{ |c| c| c c|}
 \hline
 $$&$$&$$&$$\\
${\cal S}^{12} $&$symbol$&${}^{I}\hat{\cal A}^{m \dagger}_f=$
&$\hat{b}^{m' \dagger}_{f `} *_A (\hat{b}^{m'' \dagger}_{f `})^{\dagger}$\\
\\
\hline
%
$$&$$&$$&$$\\
$1$&$\star \star$&${}^{I}\hat{\cal A}^{1 \dagger}_1=$&
$\hat{b}^{1 \dagger}_{1} *_{A} (\hat{b}^{4 \dagger}_{1})^{\dagger}$\\
$$&$$&$$&$$\\
$ $&$$ &$\stackrel{03}{[+i]}\,\stackrel{12}{(+)} \stackrel{56}{(+)}$&
$\stackrel{03}{(+i)}\,\stackrel{12}{[+]} \stackrel{56}{[+]} *_{A} 
\stackrel{03}{(-i)}\,\stackrel{12}{(+)} \stackrel{56}{(+)}$\\
\hline
$$&$$&$$&$$\\
$1$&$\ddagger$&${}^{I}\hat{\cal A}^{3 \dagger}_1=$&
$\hat{b}^{3 \dagger}_{1} *_{A} (\hat{b}^{4 \dagger}_{1})^{\dagger}$\\
$$&$$&$$&$$\\
$ $&$$ &$\stackrel{03}{(-i)}\,\stackrel{12}{(+)} \stackrel{56}{[-]}$&
$\stackrel{03}{[-i]}\,\stackrel{12}{[+]} \stackrel{56}{(-)} *_{A} 
\stackrel{03}{(-i)}\,\stackrel{12}{(+)} \stackrel{56}{(+)}$\\
\hline
$$&$$&$$&$$\\
$1$&$\odot \odot$&${}^{I}\hat{\cal A}^{1 \dagger}_4=$&
$\hat{b}^{1 \dagger}_{1} *_{A} (\hat{b}^{2 \dagger}_{1})^{\dagger}$\\
$$&$$&$$&$$\\
$ $&$$ &$\stackrel{03}{(+i)}\,\stackrel{12}{(+)} \stackrel{56}{[+]}$&
$\stackrel{03}{(+i)}\,\stackrel{12}{[+]} \stackrel{56}{[+]} *_{A} 
\stackrel{03}{[-i]}\,\stackrel{12}{(+)} \stackrel{56}{[+]}$\\
\hline
$$&$$&$$&$$\\
$1$&$\otimes$&${}^{I}\hat{\cal A}^{3 \dagger}_4=$&
$\hat{b}^{3 \dagger}_{1} *_{A} (\hat{b}^{2 \dagger}_{1})^{\dagger}$\\
$$&$$&$$&$$\\
$ $&$$ &$\stackrel{03}{[-i]}\,\stackrel{12}{(+)} \stackrel{56}{(-)}$&
$\stackrel{03}{[-i]}\,\stackrel{12}{[+]} \stackrel{56}{(-)} *_{A} 
\stackrel{03}{[-i]}\,\stackrel{12}{(+)} \stackrel{56}{[+]}$\\
\hline
\hline
$$&$$&$$&$$\\
$-1$&$\otimes$&${}^{I}\hat{\cal A}^{2 \dagger}_2=$&
$\hat{b}^{2 \dagger}_{1} *_{A} (\hat{b}^{3 \dagger}_{1})^{\dagger}$\\
$$&$$&$$&$$\\
$ $&$$ &$\stackrel{03}{[-i]}\,\stackrel{12}{(-)} \stackrel{56}{(+)}$&
$\stackrel{03}{[-i]}\,\stackrel{12}{(-)} \stackrel{56}{[+]} *_{A} 
\stackrel{03}{[-i]}\,\stackrel{12}{[+]} \stackrel{56}{(+)}$\\
\hline
$$&$$&$$&$$\\
$-1$&$\ddagger$&${}^{I}\hat{\cal A}^{4 \dagger}_2=$&
$\hat{b}^{4 \dagger}_{1} *_{A} (\hat{b}^{3 \dagger}_{1})^{\dagger}$\\
$$&$$&$$&$$\\
$ $&$$ &$\stackrel{03}{(+i)}\,\stackrel{12}{(-)} \stackrel{56}{[-]}$&
$\stackrel{03}{(+i)}\,\stackrel{12}{(-)} \stackrel{56}{(-)} *_{A} 
\stackrel{03}{[-i]}\,\stackrel{12}{[+]} \stackrel{56}{(+)}$\\
\hline
$$&$$&$$&$$\\
$-1$&$\odot \odot$&${}^{I}\hat{\cal A}^{2 \dagger}_3=$&
$\hat{b}^{2 \dagger}_{1} *_{A} (\hat{b}^{1 \dagger}_{1})^{\dagger}$\\
$$&$$&$$&$$\\
$ $&$$ &$\stackrel{03}{(-i)}\,\stackrel{12}{(-)} \stackrel{56}{[+]}$&
$\stackrel{03}{[-i]}\,\stackrel{12}{(-)} \stackrel{56}{[+]} *_{A} 
\stackrel{03}{(-i)}\,\stackrel{12}{[+]} \stackrel{56}{[+]}$\\
\hline
$$&$$&$$&$$\\
$-1$&$\star \star$&${}^{I}\hat{\cal A}^{4 \dagger}_3=$&
$\hat{b}^{4 \dagger}_{1} *_{A} (\hat{b}^{1 \dagger}_{1})^{\dagger}$\\
$$&$$&$$&$$\\
$ $&$$ &$\stackrel{03}{[+i]}\,\stackrel{12}{(-)} \stackrel{56}{(-)}$&
$\stackrel{03}{(+i)}\,\stackrel{12}{(-)} \stackrel{56}{(-)} *_{A} 
\stackrel{03}{(-i)}\,\stackrel{12}{[+]} \stackrel{56}{[+]}$\\
\hline
 \end{tabular}
 \end{center}
\end{tiny}
\end{table}
%


%
\section{One family representation of Clifford odd ``basis vectors'' in $d=(13+1)$
}
\label{13+1representation}  

This appendix, is following App.~D of Ref.~\cite{n2024NPB}.
In the even dimensional space $d=(13+1)$~(\cite{n2022epjc}, 
App.~A), one irreducible representation of the odd ``basis 
vectors'' if analysed from the point of view of the subgroups of the 
group $SO(13,1)$ (including $SO(7,1)\times SO(6)$, while $SO(7,1)$
breaks into $SO(3,1) \times SO(4)$, and $SO(6)$ breaks into 
$SU(3)\times U(1)$) contains the odd ``basis vectors'' 
describing internal spaces of quarks and leptons and antiquarks and 
antileptons, manifesting at low energies the quantum numbers 
assumed by the {\it standard model} before the electroweak break. 
Since $SO(4)$ contains two $SU(2)$ subgroups, $SU(2)_I$ and 
$SU(2)_{II}$, with the hypercharge of the {\it standard model} 
$Y=\tau^{23} + \tau^4$ (with $\tau^{23}$ belonging to 
$SU(2)_{II}$ and $\tau^4$ originating in $SO(6)$, breaking to 
$SU(3)\times U(1)$), 
one irreducible representation includes the right-handed neutrinos 
and the left-handed antineutrinos, which are not in the {\it standard 
model} scheme~\footnote{
  \begin{small}
The handedness is defined as follows
\begin{eqnarray}
\label{Gamma}
 \Gamma^{(d)}= \prod_a (\sqrt{\eta^{aa}} \gamma^a)  \cdot \left \{ \begin{array}{l l}
 (i)^{\frac{d}{2}} \,, &\rm{ for\, d \,even}\,,\\
 (i)^{\frac{d-1}{2}}\,,&\rm{for \, d \,odd}\,.
  \end{array} \right.
 \end{eqnarray}
 \end{small}
}. 
An overview of the properties of the vector and scalar gauge 
fields in the {\it spin-charge-family} theory can be found in  Refs.~(\cite{nh2021RPPNP,nd2017,n2023NPB,gmdn2007,%
gn2013} and the references therein). The reader can find in 
Table~\ref{Table so13+1.} that ``basis vectors'' of quarks have 
identical content of $SO(7,1)$ as ``basis vectors'' of leptons (and 
antiquarks as antiletons); they differ only in the 
$SU(3)\times U(1)$ part.

%
%

\bottomcaption{\label{Table so13+1.}%
\begin{small}
The left-handed ($\Gamma^{(13,1)} = -1$, Eq.~(\ref{Gamma})) irreducible representation
of one family of spinors --- the product of the odd number of nilpotents and of projectors,
which are eigenvectors of the Cartan subalgebra of the $SO(13,1)$ 
group~\cite{n2014matterantimatter,nh02}, manifesting the subgroup $SO(7,1)$ of the
colour charged quarks and antiquarks and the colourless leptons and antileptons ---
is presented.
It contains the left-handed ($\Gamma^{(3,1)}=-1$) weak ($SU(2)_{I}$) charged
($\tau^{13}=\pm \frac{1}{2}$), 
and $SU(2)_{II}$ chargeless ($\tau^{23}=0$) 
quarks and leptons, and the right-handed
($\Gamma^{(3,1)}=1$) weak ($SU(2)_{I}$) chargeless and $SU(2)_{II}$ charged
($\tau^{23}=\pm \frac{1}{2}$) quarks and leptons, both with the spin $ S^{12}$ up
and down ($\pm \frac{1}{2}$, respectively).
Quarks distinguish from leptons only in the $SU(3) \times U(1)$ part: Quarks are triplets
of three colours ($c^i$ $= (\tau^{33}, \tau^{38})$ $ = [(\frac{1}{2},\frac{1}{2\sqrt{3}}),
(-\frac{1}{2},\frac{1}{2\sqrt{3}}), (0,-\frac{1}{\sqrt{3}}) $, 
carrying the "fermion charge" ($\tau^{4}=\frac{1}{6}$). 
The colourless leptons carry the "fermion charge" ($\tau^{4}=-\frac{1}{2}$).
The same multiplet contains also the left handed weak ($SU(2)_{I}$) chargeless and
$SU(2)_{II}$ charged antiquarks and antileptons and the right handed weak
($SU(2)_{I}$) charged and $SU(2)_{II}$ chargeless antiquarks and antileptons.
Antiquarks distinguish from antileptons again only in the $SU(3) \times U(1)$ part:
Antiquarks are anti-triplets carrying the "fermion charge" ($\tau^{4}=-\frac{1}{6}$).
The anti-colourless antileptons carry the "fermion charge" ($\tau^{4}=\frac{1}{2}$).
$Y=(\tau^{23} + \tau^{4})$ is the hyper charge, the electromagnetic charge
is $Q=(\tau^{13} + Y$).
%
\end{small}
}

\tablehead{\hline
i&$$&$|^a\psi_i>$&$\Gamma^{(3,1)}$&$ S^{12}$&
$\tau^{13}$&$\tau^{23}$&$\tau^{33}$&$\tau^{38}$&$\tau^{4}$&$Y$&$Q$\\
\hline
&& ${\rm (Anti)octet},\,\Gamma^{(7,1)} = (-1)\,1\,, \,\Gamma^{(6)} = (1)\,-1$&&&&&&&&& \\
&& ${\rm of \;(anti) quarks \;and \;(anti)leptons}$&&&&&&&&&\\
\hline\hline}
\tabletail{\hline \multicolumn{12}{r}{\emph{Continued on next page}}\\}
\tablelasttail{\hline}
\begin{tiny}
\begin{supertabular}{|r|c||c||c|c||c|c||c|c|c||r|r|}
1&$ u_{R}^{c1}$&$ \stackrel{03}{(+i)}\,\stackrel{12}{[+]}|
\stackrel{56}{[+]}\,\stackrel{78}{(+)}
||\stackrel{9 \;10}{(+)}\;\;\stackrel{11\;12}{[-]}\;\;\stackrel{13\;14}{[-]} $ &1&$\frac{1}{2}$&0&
$\frac{1}{2}$&$\frac{1}{2}$&$\frac{1}{2\,\sqrt{3}}$&$\frac{1}{6}$&$\frac{2}{3}$&$\frac{2}{3}$\\
\hline
2&$u_{R}^{c1}$&$\stackrel{03}{[-i]}\,\stackrel{12}{(-)}|\stackrel{56}{[+]}\,\stackrel{78}{(+)}
||\stackrel{9 \;10}{(+)}\;\;\stackrel{11\;12}{[-]}\;\;\stackrel{13\;14}{[-]}$&1&$-\frac{1}{2}$&0&
$\frac{1}{2}$&$\frac{1}{2}$&$\frac{1}{2\,\sqrt{3}}$&$\frac{1}{6}$&$\frac{2}{3}$&$\frac{2}{3}$\\
\hline
3&$d_{R}^{c1}$&$\stackrel{03}{(+i)}\,\stackrel{12}{[+]}|\stackrel{56}{(-)}\,\stackrel{78}{[-]}
||\stackrel{9 \;10}{(+)}\;\;\stackrel{11\;12}{[-]}\;\;\stackrel{13\;14}{[-]}$&1&$\frac{1}{2}$&0&
$-\frac{1}{2}$&$\frac{1}{2}$&$\frac{1}{2\,\sqrt{3}}$&$\frac{1}{6}$&$-\frac{1}{3}$&$-\frac{1}{3}$\\
\hline
4&$ d_{R}^{c1} $&$\stackrel{03}{[-i]}\,\stackrel{12}{(-)}|
\stackrel{56}{(-)}\,\stackrel{78}{[-]}
||\stackrel{9 \;10}{(+)}\;\;\stackrel{11\;12}{[-]}\;\;\stackrel{13\;14}{[-]} $&1&$-\frac{1}{2}$&0&
$-\frac{1}{2}$&$\frac{1}{2}$&$\frac{1}{2\,\sqrt{3}}$&$\frac{1}{6}$&$-\frac{1}{3}$&$-\frac{1}{3}$\\
\hline
5&$d_{L}^{c1}$&$\stackrel{03}{[-i]}\,\stackrel{12}{[+]}|\stackrel{56}{(-)}\,\stackrel{78}{(+)}
||\stackrel{9 \;10}{(+)}\;\;\stackrel{11\;12}{[-]}\;\;\stackrel{13\;14}{[-]}$&-1&$\frac{1}{2}$&
$-\frac{1}{2}$&0&$\frac{1}{2}$&$\frac{1}{2\,\sqrt{3}}$&$\frac{1}{6}$&$\frac{1}{6}$&$-\frac{1}{3}$\\
\hline
6&$d_{L}^{c1} $&$ - \stackrel{03}{(+i)}\,\stackrel{12}{(-)}|\stackrel{56}{(-)}\,\stackrel{78}{(+)}
||\stackrel{9 \;10}{(+)}\;\;\stackrel{11\;12}{[-]}\;\;\stackrel{13\;14}{[-]} $&-1&$-\frac{1}{2}$&
$-\frac{1}{2}$&0&$\frac{1}{2}$&$\frac{1}{2\,\sqrt{3}}$&$\frac{1}{6}$&$\frac{1}{6}$&$-\frac{1}{3}$\\
\hline
7&$ u_{L}^{c1}$&$ - \stackrel{03}{[-i]}\,\stackrel{12}{[+]}|\stackrel{56}{[+]}\,\stackrel{78}{[-]}
||\stackrel{9 \;10}{(+)}\;\;\stackrel{11\;12}{[-]}\;\;\stackrel{13\;14}{[-]}$ &-1&$\frac{1}{2}$&
$\frac{1}{2}$&0 &$\frac{1}{2}$&$\frac{1}{2\,\sqrt{3}}$&$\frac{1}{6}$&$\frac{1}{6}$&$\frac{2}{3}$\\
\hline
8&$u_{L}^{c1}$&$\stackrel{03}{(+i)}\,\stackrel{12}{(-)}|\stackrel{56}{[+]}\,\stackrel{78}{[-]}
||\stackrel{9 \;10}{(+)}\;\;\stackrel{11\;12}{[-]}\;\;\stackrel{13\;14}{[-]}$&-1&$-\frac{1}{2}$&
$\frac{1}{2}$&0&$\frac{1}{2}$&$\frac{1}{2\,\sqrt{3}}$&$\frac{1}{6}$&$\frac{1}{6}$&$\frac{2}{3}$\\
\hline\hline
\shrinkheight{0.25\textheight}
9&$ u_{R}^{c2}$&$ \stackrel{03}{(+i)}\,\stackrel{12}{[+]}|
\stackrel{56}{[+]}\,\stackrel{78}{(+)}
||\stackrel{9 \;10}{[-]}\;\;\stackrel{11\;12}{(+)}\;\;\stackrel{13\;14}{[-]} $ &1&$\frac{1}{2}$&0&
$\frac{1}{2}$&$-\frac{1}{2}$&$\frac{1}{2\,\sqrt{3}}$&$\frac{1}{6}$&$\frac{2}{3}$&$\frac{2}{3}$\\
\hline
10&$u_{R}^{c2}$&$\stackrel{03}{[-i]}\,\stackrel{12}{(-)}|\stackrel{56}{[+]}\,\stackrel{78}{(+)}
||\stackrel{9 \;10}{[-]}\;\;\stackrel{11\;12}{(+)}\;\;\stackrel{13\;14}{[-]}$&1&$-\frac{1}{2}$&0&
$\frac{1}{2}$&$-\frac{1}{2}$&$\frac{1}{2\,\sqrt{3}}$&$\frac{1}{6}$&$\frac{2}{3}$&$\frac{2}{3}$\\
\hline
11&$d_{R}^{c2}$&$\stackrel{03}{(+i)}\,\stackrel{12}{[+]}|\stackrel{56}{(-)}\,\stackrel{78}{[-]}
||\stackrel{9 \;10}{[-]}\;\;\stackrel{11\;12}{(+)}\;\;\stackrel{13\;14}{[-]}$
&1&$\frac{1}{2}$&0&
$-\frac{1}{2}$&$ - \frac{1}{2}$&$\frac{1}{2\,\sqrt{3}}$&$\frac{1}{6}$&$-\frac{1}{3}$&$-\frac{1}{3}$\\
\hline
12&$ d_{R}^{c2} $&$\stackrel{03}{[-i]}\,\stackrel{12}{(-)}|
\stackrel{56}{(-)}\,\stackrel{78}{[-]}
||\stackrel{9 \;10}{[-]}\;\;\stackrel{11\;12}{(+)}\;\;\stackrel{13\;14}{[-]} $
&1&$-\frac{1}{2}$&0&
$-\frac{1}{2}$&$-\frac{1}{2}$&$\frac{1}{2\,\sqrt{3}}$&$\frac{1}{6}$&$-\frac{1}{3}$&$-\frac{1}{3}$\\
\hline
13&$d_{L}^{c2}$&$\stackrel{03}{[-i]}\,\stackrel{12}{[+]}|\stackrel{56}{(-)}\,\stackrel{78}{(+)}
||\stackrel{9 \;10}{[-]}\;\;\stackrel{11\;12}{(+)}\;\;\stackrel{13\;14}{[-]}$
&-1&$\frac{1}{2}$&
$-\frac{1}{2}$&0&$-\frac{1}{2}$&$\frac{1}{2\,\sqrt{3}}$&$\frac{1}{6}$&$\frac{1}{6}$&$-\frac{1}{3}$\\
\hline
14&$d_{L}^{c2} $&$ - \stackrel{03}{(+i)}\,\stackrel{12}{(-)}|\stackrel{56}{(-)}\,\stackrel{78}{(+)}
||\stackrel{9 \;10}{[-]}\;\;\stackrel{11\;12}{(+)}\;\;\stackrel{13\;14}{[-]} $&-1&$-\frac{1}{2}$&
$-\frac{1}{2}$&0&$-\frac{1}{2}$&$\frac{1}{2\,\sqrt{3}}$&$\frac{1}{6}$&$\frac{1}{6}$&$-\frac{1}{3}$\\
\hline
15&$ u_{L}^{c2}$&$ - \stackrel{03}{[-i]}\,\stackrel{12}{[+]}|\stackrel{56}{[+]}\,\stackrel{78}{[-]}
||\stackrel{9 \;10}{[-]}\;\;\stackrel{11\;12}{(+)}\;\;\stackrel{13\;14}{[-]}$ &-1&$\frac{1}{2}$&
$\frac{1}{2}$&0 &$-\frac{1}{2}$&$\frac{1}{2\,\sqrt{3}}$&$\frac{1}{6}$&$\frac{1}{6}$&$\frac{2}{3}$\\
\hline
16&$u_{L}^{c2}$&$\stackrel{03}{(+i)}\,\stackrel{12}{(-)}|\stackrel{56}{[+]}\,\stackrel{78}{[-]}
||\stackrel{9 \;10}{[-]}\;\;\stackrel{11\;12}{(+)}\;\;\stackrel{13\;14}{[-]}$&-1&$-\frac{1}{2}$&
$\frac{1}{2}$&0&$-\frac{1}{2}$&$\frac{1}{2\,\sqrt{3}}$&$\frac{1}{6}$&$\frac{1}{6}$&$\frac{2}{3}$\\
\hline\hline
17&$ u_{R}^{c3}$&$ \stackrel{03}{(+i)}\,\stackrel{12}{[+]}|
\stackrel{56}{[+]}\,\stackrel{78}{(+)}
||\stackrel{9 \;10}{[-]}\;\;\stackrel{11\;12}{[-]}\;\;\stackrel{13\;14}{(+)} $ &1&$\frac{1}{2}$&0&
$\frac{1}{2}$&$0$&$-\frac{1}{\sqrt{3}}$&$\frac{1}{6}$&$\frac{2}{3}$&$\frac{2}{3}$\\
\hline
18&$u_{R}^{c3}$&$\stackrel{03}{[-i]}\,\stackrel{12}{(-)}|\stackrel{56}{[+]}\,\stackrel{78}{(+)}
||\stackrel{9 \;10}{[-]}\;\;\stackrel{11\;12}{[-]}\;\;\stackrel{13\;14}{(+)}$&1&$-\frac{1}{2}$&0&
$\frac{1}{2}$&$0$&$-\frac{1}{\sqrt{3}}$&$\frac{1}{6}$&$\frac{2}{3}$&$\frac{2}{3}$\\
\hline
19&$d_{R}^{c3}$&$\stackrel{03}{(+i)}\,\stackrel{12}{[+]}|\stackrel{56}{(-)}\,\stackrel{78}{[-]}
||\stackrel{9 \;10}{[-]}\;\;\stackrel{11\;12}{[-]}\;\;\stackrel{13\;14}{(+)}$&1&$\frac{1}{2}$&0&
$-\frac{1}{2}$&$0$&$-\frac{1}{\sqrt{3}}$&$\frac{1}{6}$&$-\frac{1}{3}$&$-\frac{1}{3}$\\
\hline
20&$ d_{R}^{c3} $&$\stackrel{03}{[-i]}\,\stackrel{12}{(-)}|
\stackrel{56}{(-)}\,\stackrel{78}{[-]}
||\stackrel{9 \;10}{[-]}\;\;\stackrel{11\;12}{[-]}\;\;\stackrel{13\;14}{(+)} $&1&$-\frac{1}{2}$&0&
$-\frac{1}{2}$&$0$&$-\frac{1}{\sqrt{3}}$&$\frac{1}{6}$&$-\frac{1}{3}$&$-\frac{1}{3}$\\
\hline
21&$d_{L}^{c3}$&$\stackrel{03}{[-i]}\,\stackrel{12}{[+]}|\stackrel{56}{(-)}\,\stackrel{78}{(+)}
||\stackrel{9 \;10}{[-]}\;\;\stackrel{11\;12}{[-]}\;\;\stackrel{13\;14}{(+)}$&-1&$\frac{1}{2}$&
$-\frac{1}{2}$&0&$0$&$-\frac{1}{\sqrt{3}}$&$\frac{1}{6}$&$\frac{1}{6}$&$-\frac{1}{3}$\\
\hline
22&$d_{L}^{c3} $&$ - \stackrel{03}{(+i)}\,\stackrel{12}{(-)}|\stackrel{56}{(-)}\,\stackrel{78}{(+)}
||\stackrel{9 \;10}{[-]}\;\;\stackrel{11\;12}{[-]}\;\;\stackrel{13\;14}{(+)} $&-1&$-\frac{1}{2}$&
$-\frac{1}{2}$&0&$0$&$-\frac{1}{\sqrt{3}}$&$\frac{1}{6}$&$\frac{1}{6}$&$-\frac{1}{3}$\\
\hline
23&$ u_{L}^{c3}$&$ - \stackrel{03}{[-i]}\,\stackrel{12}{[+]}|\stackrel{56}{[+]}\,\stackrel{78}{[-]}
||\stackrel{9 \;10}{[-]}\;\;\stackrel{11\;12}{[-]}\;\;\stackrel{13\;14}{(+)}$ &-1&$\frac{1}{2}$&
$\frac{1}{2}$&0 &$0$&$-\frac{1}{\sqrt{3}}$&$\frac{1}{6}$&$\frac{1}{6}$&$\frac{2}{3}$\\
\hline
24&$u_{L}^{c3}$&$\stackrel{03}{(+i)}\,\stackrel{12}{(-)}|\stackrel{56}{[+]}\,\stackrel{78}{[-]}
||\stackrel{9 \;10}{[-]}\;\;\stackrel{11\;12}{[-]}\;\;\stackrel{13\;14}{(+)}$&-1&$-\frac{1}{2}$&
$\frac{1}{2}$&0&$0$&$-\frac{1}{\sqrt{3}}$&$\frac{1}{6}$&$\frac{1}{6}$&$\frac{2}{3}$\\
\hline\hline
25&$ \nu_{R}$&$ \stackrel{03}{(+i)}\,\stackrel{12}{[+]}|
\stackrel{56}{[+]}\,\stackrel{78}{(+)}
||\stackrel{9 \;10}{(+)}\;\;\stackrel{11\;12}{(+)}\;\;\stackrel{13\;14}{(+)} $ &1&$\frac{1}{2}$&0&
$\frac{1}{2}$&$0$&$0$&$-\frac{1}{2}$&$0$&$0$\\
\hline
26&$\nu_{R}$&$\stackrel{03}{[-i]}\,\stackrel{12}{(-)}|\stackrel{56}{[+]}\,\stackrel{78}{(+)}
||\stackrel{9 \;10}{(+)}\;\;\stackrel{11\;12}{(+)}\;\;\stackrel{13\;14}{(+)}$&1&$-\frac{1}{2}$&0&
$\frac{1}{2}$ &$0$&$0$&$-\frac{1}{2}$&$0$&$0$\\
\hline
27&$e_{R}$&$\stackrel{03}{(+i)}\,\stackrel{12}{[+]}|\stackrel{56}{(-)}\,\stackrel{78}{[-]}
||\stackrel{9 \;10}{(+)}\;\;\stackrel{11\;12}{(+)}\;\;\stackrel{13\;14}{(+)}$&1&$\frac{1}{2}$&0&
$-\frac{1}{2}$&$0$&$0$&$-\frac{1}{2}$&$-1$&$-1$\\
\hline
28&$ e_{R} $&$\stackrel{03}{[-i]}\,\stackrel{12}{(-)}|
\stackrel{56}{(-)}\,\stackrel{78}{[-]}
||\stackrel{9 \;10}{(+)}\;\;\stackrel{11\;12}{(+)}\;\;\stackrel{13\;14}{(+)} $&1&$-\frac{1}{2}$&0&
$-\frac{1}{2}$&$0$&$0$&$-\frac{1}{2}$&$-1$&$-1$\\
\hline
29&$e_{L}$&$\stackrel{03}{[-i]}\,\stackrel{12}{[+]}|\stackrel{56}{(-)}\,\stackrel{78}{(+)}
||\stackrel{9 \;10}{(+)}\;\;\stackrel{11\;12}{(+)}\;\;\stackrel{13\;14}{(+)}$&-1&$\frac{1}{2}$&
$-\frac{1}{2}$&0&$0$&$0$&$-\frac{1}{2}$&$-\frac{1}{2}$&$-1$\\
\hline
30&$e_{L} $&$ - \stackrel{03}{(+i)}\,\stackrel{12}{(-)}|\stackrel{56}{(-)}\,\stackrel{78}{(+)}
||\stackrel{9 \;10}{(+)}\;\;\stackrel{11\;12}{(+)}\;\;\stackrel{13\;14}{(+)} $&-1&$-\frac{1}{2}$&
$-\frac{1}{2}$&0&$0$&$0$&$-\frac{1}{2}$&$-\frac{1}{2}$&$-1$\\
\hline
31&$ \nu_{L}$&$ - \stackrel{03}{[-i]}\,\stackrel{12}{[+]}|\stackrel{56}{[+]}\,\stackrel{78}{[-]}
||\stackrel{9 \;10}{(+)}\;\;\stackrel{11\;12}{(+)}\;\;\stackrel{13\;14}{(+)}$ &-1&$\frac{1}{2}$&
$\frac{1}{2}$&0 &$0$&$0$&$-\frac{1}{2}$&$-\frac{1}{2}$&$0$\\
\hline
32&$\nu_{L}$&$\stackrel{03}{(+i)}\,\stackrel{12}{(-)}|\stackrel{56}{[+]}\,\stackrel{78}{[-]}
||\stackrel{9 \;10}{(+)}\;\;\stackrel{11\;12}{(+)}\;\;\stackrel{13\;14}{(+)}$&-1&$-\frac{1}{2}$&
$\frac{1}{2}$&0&$0$&$0$&$-\frac{1}{2}$&$-\frac{1}{2}$&$0$\\
\hline\hline
33&$ \bar{d}_{L}^{\bar{c1}}$&$ \stackrel{03}{[-i]}\,\stackrel{12}{[+]}|
\stackrel{56}{[+]}\,\stackrel{78}{(+)}
||\stackrel{9 \;10}{[-]}\;\;\stackrel{11\;12}{(+)}\;\;\stackrel{13\;14}{(+)} $ &-1&$\frac{1}{2}$&0&
$\frac{1}{2}$&$-\frac{1}{2}$&$-\frac{1}{2\,\sqrt{3}}$&$-\frac{1}{6}$&$\frac{1}{3}$&$\frac{1}{3}$\\
\hline
34&$\bar{d}_{L}^{\bar{c1}}$&$\stackrel{03}{(+i)}\,\stackrel{12}{(-)}|\stackrel{56}{[+]}\,\stackrel{78}{(+)}
||\stackrel{9 \;10}{[-]}\;\;\stackrel{11\;12}{(+)}\;\;\stackrel{13\;14}{(+)}$&-1&$-\frac{1}{2}$&0&
$\frac{1}{2}$&$-\frac{1}{2}$&$-\frac{1}{2\,\sqrt{3}}$&$-\frac{1}{6}$&$\frac{1}{3}$&$\frac{1}{3}$\\
\hline
35&$\bar{u}_{L}^{\bar{c1}}$&$ - \stackrel{03}{[-i]}\,\stackrel{12}{[+]}|\stackrel{56}{(-)}\,\stackrel{78}{[-]}
||\stackrel{9 \;10}{[-]}\;\;\stackrel{11\;12}{(+)}\;\;\stackrel{13\;14}{(+)}$&-1&$\frac{1}{2}$&0&
$-\frac{1}{2}$&$-\frac{1}{2}$&$-\frac{1}{2\,\sqrt{3}}$&$-\frac{1}{6}$&$-\frac{2}{3}$&$-\frac{2}{3}$\\
\hline
36&$ \bar{u}_{L}^{\bar{c1}} $&$ - \stackrel{03}{(+i)}\,\stackrel{12}{(-)}|
\stackrel{56}{(-)}\,\stackrel{78}{[-]}
||\stackrel{9 \;10}{[-]}\;\;\stackrel{11\;12}{(+)}\;\;\stackrel{13\;14}{(+)} $&-1&$-\frac{1}{2}$&0&
$-\frac{1}{2}$&$-\frac{1}{2}$&$-\frac{1}{2\,\sqrt{3}}$&$-\frac{1}{6}$&$-\frac{2}{3}$&$-\frac{2}{3}$\\
\hline
37&$\bar{d}_{R}^{\bar{c1}}$&$\stackrel{03}{(+i)}\,\stackrel{12}{[+]}|\stackrel{56}{[+]}\,\stackrel{78}{[-]}
||\stackrel{9 \;10}{[-]}\;\;\stackrel{11\;12}{(+)}\;\;\stackrel{13\;14}{(+)}$&1&$\frac{1}{2}$&
$\frac{1}{2}$&0&$-\frac{1}{2}$&$-\frac{1}{2\,\sqrt{3}}$&$-\frac{1}{6}$&$-\frac{1}{6}$&$\frac{1}{3}$\\
\hline
38&$\bar{d}_{R}^{\bar{c1}} $&$ - \stackrel{03}{[-i]}\,\stackrel{12}{(-)}|\stackrel{56}{[+]}\,\stackrel{78}{[-]}
||\stackrel{9 \;10}{[-]}\;\;\stackrel{11\;12}{(+)}\;\;\stackrel{13\;14}{(+)} $&1&$-\frac{1}{2}$&
$\frac{1}{2}$&0&$-\frac{1}{2}$&$-\frac{1}{2\,\sqrt{3}}$&$-\frac{1}{6}$&$-\frac{1}{6}$&$\frac{1}{3}$\\
\hline
39&$ \bar{u}_{R}^{\bar{c1}}$&$\stackrel{03}{(+i)}\,\stackrel{12}{[+]}|\stackrel{56}{(-)}\,\stackrel{78}{(+)}
||\stackrel{9 \;10}{[-]}\;\;\stackrel{11\;12}{(+)}\;\;\stackrel{13\;14}{(+)}$ &1&$\frac{1}{2}$&
$-\frac{1}{2}$&0 &$-\frac{1}{2}$&$-\frac{1}{2\,\sqrt{3}}$&$-\frac{1}{6}$&$-\frac{1}{6}$&$-\frac{2}{3}$\\
\hline
40&$\bar{u}_{R}^{\bar{c1}}$&$\stackrel{03}{[-i]}\,\stackrel{12}{(-)}|\stackrel{56}{(-)}\,\stackrel{78}{(+)}
||\stackrel{9 \;10}{[-]}\;\;\stackrel{11\;12}{(+)}\;\;\stackrel{13\;14}{(+)}$
&1&$-\frac{1}{2}$&
$-\frac{1}{2}$&0&$-\frac{1}{2}$&$-\frac{1}{2\,\sqrt{3}}$&$-\frac{1}{6}$&$-\frac{1}{6}$&$-\frac{2}{3}$\\
\hline\hline
41&$ \bar{d}_{L}^{\bar{c2}}$&$ \stackrel{03}{[-i]}\,\stackrel{12}{[+]}|
\stackrel{56}{[+]}\,\stackrel{78}{(+)}
||\stackrel{9 \;10}{(+)}\;\;\stackrel{11\;12}{[-]}\;\;\stackrel{13\;14}{(+)} $
&-1&$\frac{1}{2}$&0&
$\frac{1}{2}$&$\frac{1}{2}$&$-\frac{1}{2\,\sqrt{3}}$&$-\frac{1}{6}$&$\frac{1}{3}$&$\frac{1}{3}$\\
\hline
42&$\bar{d}_{L}^{\bar{c2}}$&$\stackrel{03}{(+i)}\,\stackrel{12}{(-)}|\stackrel{56}{[+]}\,\stackrel{78}{(+)}
||\stackrel{9 \;10}{(+)}\;\;\stackrel{11\;12}{[-]}\;\;\stackrel{13\;14}{(+)}$
&-1&$-\frac{1}{2}$&0&
$\frac{1}{2}$&$\frac{1}{2}$&$-\frac{1}{2\,\sqrt{3}}$&$-\frac{1}{6}$&$\frac{1}{3}$&$\frac{1}{3}$\\
\hline
43&$\bar{u}_{L}^{\bar{c2}}$&$ - \stackrel{03}{[-i]}\,\stackrel{12}{[+]}|\stackrel{56}{(-)}\,\stackrel{78}{[-]}
||\stackrel{9 \;10}{(+)}\;\;\stackrel{11\;12}{[-]}\;\;\stackrel{13\;14}{(+)}$
&-1&$\frac{1}{2}$&0&
$-\frac{1}{2}$&$\frac{1}{2}$&$-\frac{1}{2\,\sqrt{3}}$&$-\frac{1}{6}$&$-\frac{2}{3}$&$-\frac{2}{3}$\\
\hline
44&$ \bar{u}_{L}^{\bar{c2}} $&$ - \stackrel{03}{(+i)}\,\stackrel{12}{(-)}|
\stackrel{56}{(-)}\,\stackrel{78}{[-]}
||\stackrel{9 \;10}{(+)}\;\;\stackrel{11\;12}{[-]}\;\;\stackrel{13\;14}{(+)} $
&-1&$-\frac{1}{2}$&0&
$-\frac{1}{2}$&$\frac{1}{2}$&$-\frac{1}{2\,\sqrt{3}}$&$-\frac{1}{6}$&$-\frac{2}{3}$&$-\frac{2}{3}$\\
\hline
45&$\bar{d}_{R}^{\bar{c2}}$&$\stackrel{03}{(+i)}\,\stackrel{12}{[+]}|\stackrel{56}{[+]}\,\stackrel{78}{[-]}
||\stackrel{9 \;10}{(+)}\;\;\stackrel{11\;12}{[-]}\;\;\stackrel{13\;14}{(+)}$
&1&$\frac{1}{2}$&
$\frac{1}{2}$&0&$\frac{1}{2}$&$-\frac{1}{2\,\sqrt{3}}$&$-\frac{1}{6}$&$-\frac{1}{6}$&$\frac{1}{3}$\\
\hline
46&$\bar{d}_{R}^{\bar{c2}} $&$ - \stackrel{03}{[-i]}\,\stackrel{12}{(-)}|\stackrel{56}{[+]}\,\stackrel{78}{[-]}
||\stackrel{9 \;10}{(+)}\;\;\stackrel{11\;12}{[-]}\;\;\stackrel{13\;14}{(+)} $
&1&$-\frac{1}{2}$&
$\frac{1}{2}$&0&$\frac{1}{2}$&$-\frac{1}{2\,\sqrt{3}}$&$-\frac{1}{6}$&$-\frac{1}{6}$&$\frac{1}{3}$\\
\hline
47&$ \bar{u}_{R}^{\bar{c2}}$&$\stackrel{03}{(+i)}\,\stackrel{12}{[+]}|\stackrel{56}{(-)}\,\stackrel{78}{(+)}
||\stackrel{9 \;10}{(+)}\;\;\stackrel{11\;12}{[-]}\;\;\stackrel{13\;14}{(+)}$
 &1&$\frac{1}{2}$&
$-\frac{1}{2}$&0 &$\frac{1}{2}$&$-\frac{1}{2\,\sqrt{3}}$&$-\frac{1}{6}$&$-\frac{1}{6}$&$-\frac{2}{3}$\\
\hline
48&$\bar{u}_{R}^{\bar{c2}}$&$\stackrel{03}{[-i]}\,\stackrel{12}{(-)}|\stackrel{56}{(-)}\,\stackrel{78}{(+)}
||\stackrel{9 \;10}{(+)}\;\;\stackrel{11\;12}{[-]}\;\;\stackrel{13\;14}{(+)}$
&1&$-\frac{1}{2}$&
$-\frac{1}{2}$&0&$\frac{1}{2}$&$-\frac{1}{2\,\sqrt{3}}$&$-\frac{1}{6}$&$-\frac{1}{6}$&$-\frac{2}{3}$\\
\hline\hline
49&$ \bar{d}_{L}^{\bar{c3}}$&$ \stackrel{03}{[-i]}\,\stackrel{12}{[+]}|
\stackrel{56}{[+]}\,\stackrel{78}{(+)}
||\stackrel{9 \;10}{(+)}\;\;\stackrel{11\;12}{(+)}\;\;\stackrel{13\;14}{[-]} $ &-1&$\frac{1}{2}$&0&
$\frac{1}{2}$&$0$&$\frac{1}{\sqrt{3}}$&$-\frac{1}{6}$&$\frac{1}{3}$&$\frac{1}{3}$\\
\hline
50&$\bar{d}_{L}^{\bar{c3}}$&$\stackrel{03}{(+i)}\,\stackrel{12}{(-)}|\stackrel{56}{[+]}\,\stackrel{78}{(+)}
||\stackrel{9 \;10}{(+)}\;\;\stackrel{11\;12}{(+)}\;\;\stackrel{13\;14}{[-]} $&-1&$-\frac{1}{2}$&0&
$\frac{1}{2}$&$0$&$\frac{1}{\sqrt{3}}$&$-\frac{1}{6}$&$\frac{1}{3}$&$\frac{1}{3}$\\
\hline
51&$\bar{u}_{L}^{\bar{c3}}$&$ - \stackrel{03}{[-i]}\,\stackrel{12}{[+]}|\stackrel{56}{(-)}\,\stackrel{78}{[-]}
||\stackrel{9 \;10}{(+)}\;\;\stackrel{11\;12}{(+)}\;\;\stackrel{13\;14}{[-]} $&-1&$\frac{1}{2}$&0&
$-\frac{1}{2}$&$0$&$\frac{1}{\sqrt{3}}$&$-\frac{1}{6}$&$-\frac{2}{3}$&$-\frac{2}{3}$\\
\hline
52&$ \bar{u}_{L}^{\bar{c3}} $&$ - \stackrel{03}{(+i)}\,\stackrel{12}{(-)}|
\stackrel{56}{(-)}\,\stackrel{78}{[-]}
||\stackrel{9 \;10}{(+)}\;\;\stackrel{11\;12}{(+)}\;\;\stackrel{13\;14}{[-]}  $&-1&$-\frac{1}{2}$&0&
$-\frac{1}{2}$&$0$&$\frac{1}{\sqrt{3}}$&$-\frac{1}{6}$&$-\frac{2}{3}$&$-\frac{2}{3}$\\
\hline
53&$\bar{d}_{R}^{\bar{c3}}$&$\stackrel{03}{(+i)}\,\stackrel{12}{[+]}|\stackrel{56}{[+]}\,\stackrel{78}{[-]}
||\stackrel{9 \;10}{(+)}\;\;\stackrel{11\;12}{(+)}\;\;\stackrel{13\;14}{[-]} $&1&$\frac{1}{2}$&
$\frac{1}{2}$&0&$0$&$\frac{1}{\sqrt{3}}$&$-\frac{1}{6}$&$-\frac{1}{6}$&$\frac{1}{3}$\\
\hline
54&$\bar{d}_{R}^{\bar{c3}} $&$ - \stackrel{03}{[-i]}\,\stackrel{12}{(-)}|\stackrel{56}{[+]}\,\stackrel{78}{[-]}
||\stackrel{9 \;10}{(+)}\;\;\stackrel{11\;12}{(+)}\;\;\stackrel{13\;14}{[-]} $&1&$-\frac{1}{2}$&
$\frac{1}{2}$&0&$0$&$\frac{1}{\sqrt{3}}$&$-\frac{1}{6}$&$-\frac{1}{6}$&$\frac{1}{3}$\\
\hline
55&$ \bar{u}_{R}^{\bar{c3}}$&$\stackrel{03}{(+i)}\,\stackrel{12}{[+]}|\stackrel{56}{(-)}\,\stackrel{78}{(+)}
||\stackrel{9 \;10}{(+)}\;\;\stackrel{11\;12}{(+)}\;\;\stackrel{13\;14}{[-]} $ &1&$\frac{1}{2}$&
$-\frac{1}{2}$&0 &$0$&$\frac{1}{\sqrt{3}}$&$-\frac{1}{6}$&$-\frac{1}{6}$&$-\frac{2}{3}$\\
\hline
56&$\bar{u}_{R}^{\bar{c3}}$&$\stackrel{03}{[-i]}\,\stackrel{12}{(-)}|\stackrel{56}{(-)}\,\stackrel{78}{(+)}
||\stackrel{9 \;10}{(+)}\;\;\stackrel{11\;12}{(+)}\;\;\stackrel{13\;14}{[-]} $&1&$-\frac{1}{2}$&
$-\frac{1}{2}$&0&$0$&$\frac{1}{\sqrt{3}}$&$-\frac{1}{6}$&$-\frac{1}{6}$&$-\frac{2}{3}$\\
\hline\hline
57&$ \bar{e}_{L}$&$ \stackrel{03}{[-i]}\,\stackrel{12}{[+]}|
\stackrel{56}{[+]}\,\stackrel{78}{(+)}
||\stackrel{9 \;10}{[-]}\;\;\stackrel{11\;12}{[-]}\;\;\stackrel{13\;14}{[-]} $ &-1&$\frac{1}{2}$&0&
$\frac{1}{2}$&$0$&$0$&$\frac{1}{2}$&$1$&$1$\\
\hline
58&$\bar{e}_{L}$&$\stackrel{03}{(+i)}\,\stackrel{12}{(-)}|\stackrel{56}{[+]}\,\stackrel{78}{(+)}
||\stackrel{9 \;10}{[-]}\;\;\stackrel{11\;12}{[-]}\;\;\stackrel{13\;14}{[-]}$&-1&$-\frac{1}{2}$&0&
$\frac{1}{2}$ &$0$&$0$&$\frac{1}{2}$&$1$&$1$\\
\hline
59&$\bar{\nu}_{L}$&$ - \stackrel{03}{[-i]}\,\stackrel{12}{[+]}|\stackrel{56}{(-)}\,\stackrel{78}{[-]}
||\stackrel{9 \;10}{[-]}\;\;\stackrel{11\;12}{[-]}\;\;\stackrel{13\;14}{[-]}$&-1&$\frac{1}{2}$&0&
$-\frac{1}{2}$&$0$&$0$&$\frac{1}{2}$&$0$&$0$\\
\hline
60&$ \bar{\nu}_{L} $&$ - \stackrel{03}{(+i)}\,\stackrel{12}{(-)}|
\stackrel{56}{(-)}\,\stackrel{78}{[-]}
||\stackrel{9 \;10}{[-]}\;\;\stackrel{11\;12}{[-]}\;\;\stackrel{13\;14}{[-]} $&-1&$-\frac{1}{2}$&0&
$-\frac{1}{2}$&$0$&$0$&$\frac{1}{2}$&$0$&$0$\\
\hline
61&$\bar{\nu}_{R}$&$\stackrel{03}{(+i)}\,\stackrel{12}{[+]}|\stackrel{56}{(-)}\,\stackrel{78}{(+)}
||\stackrel{9 \;10}{[-]}\;\;\stackrel{11\;12}{[-]}\;\;\stackrel{13\;14}{[-]}$&1&$\frac{1}{2}$&
$-\frac{1}{2}$&0&$0$&$0$&$\frac{1}{2}$&$\frac{1}{2}$&$0$\\
\hline
62&$\bar{\nu}_{R} $&$ - \stackrel{03}{[-i]}\,\stackrel{12}{(-)}|\stackrel{56}{(-)}\,\stackrel{78}{(+)}
||\stackrel{9 \;10}{[-]}\;\;\stackrel{11\;12}{[-]}\;\;\stackrel{13\;14}{[-]} $&1&$-\frac{1}{2}$&
$-\frac{1}{2}$&0&$0$&$0$&$\frac{1}{2}$&$\frac{1}{2}$&$0$\\
\hline
63&$ \bar{e}_{R}$&$\stackrel{03}{(+i)}\,\stackrel{12}{[+]}|\stackrel{56}{[+]}\,\stackrel{78}{[-]}
||\stackrel{9 \;10}{[-]}\;\;\stackrel{11\;12}{[-]}\;\;\stackrel{13\;14}{[-]}$ &1&$\frac{1}{2}$&
$\frac{1}{2}$&0 &$0$&$0$&$\frac{1}{2}$&$\frac{1}{2}$&$1$\\
\hline
64&$\bar{e}_{R}$&$\stackrel{03}{[-i]}\,\stackrel{12}{(-)}|\stackrel{56}{[+]}\,\stackrel{78}{[-]}
||\stackrel{9 \;10}{[-]}\;\;\stackrel{11\;12}{[-]}\;\;\stackrel{13\;14}{[-]}$&1&$-\frac{1}{2}$&
$\frac{1}{2}$&0&$0$&$0$&$\frac{1}{2}$&$\frac{1}{2}$&$1$\\
\hline
\end{supertabular}
\end{tiny}

\vspace {3mm}
 
\section{Acknowledgement}
 The author N.S.M.B. thanks Department of Physics, FMF, University of Ljubljana, 
Society of  Mathematicians, Physicists and Astronomers of Slovenia for supporting  
the research on the  {\it spin-charge-family} theory, and
 Matja\v z Breskvar of  Beyond Semiconductor for donations, in particular for 
sponsoring the annual workshops entitled  "What comes beyond the standard
 models" at Bled, in which the ideas and realizations, presented in this paper,
 were discussed. The author thanks  Milutin Blagojevi\' c
  for fruitful discussions.


\begin{thebibliography}{99}
\bibitem{n2024NPB}  N. S. Manko\v c Bor\v stnik, "Can the ``basis vectors'', 
describing t he internal spaces of fermion and boson fields with the Clifford odd 
(for fermion) and Clifford even (for boson) objects, explain interactions 
among fields, with gravitons included?"  [arxiv: 2407.09482]. 
\end{thebibliography}
\end{document}